## Author's Errata and Note on Version 3

*Deep Learning Based on Generative Adversarial and Convolutional Neural Networks for Financial Time Series Predictions*

Wilfredo Tovar Hidalgo — arXiv:2008.08041 — June 2026

---

This version adds an errata note. **The body of the paper is unchanged from v2.** The corrections below were identified by the author during a systematic re-review conducted in 2026 in preparation for a follow-up study. They are recorded here so that readers and citing authors are not misled by the original text.

**1. Market coverage is overstated in the abstract.** The abstract lists TSX, SHCOMP, KOSPI 200 and the S&P 500. The data section (Sec. IV-A) names TSE, SHCOMP and the S&P 500, and the reported results (Sec. VI) cover three securities listed on the Nasdaq Composite, the Shanghai Composite and the Toronto Stock Exchange. No Korean data is described as collected and no Korean result is reported. **Readers should treat the study as covering three exchanges.**

**2. Sample size should have been stated more clearly.** The 2,954-stock figure describes the available universe. Twelve individual stock records were used for training, and the trend classification results reported in Figs. 15–17 are based on three securities. The paper does not make this relationship explicit and should have.

**3. Data frequency is stated inconsistently.** Sec. IV-A describes daily data "on a working day basis"; Sec. IV-E describes windows of 14 intervals at 10-minute resolution (140 minutes); Sec. VI refers to "a scale of variation per minute." At most one of these can describe the data actually used. The author is unable to resolve this from surviving records and flags it as an unresolved ambiguity in the original.

**4. The evaluation protocol is under-specified.** The paper does not state a train/test partition rule, does not state whether recursive feature elimination and randomized PCA were fit inside a training partition or on the full sample, and does not mention normalisation. These omissions mean the reported result is not reconstructible from the published description, and readers should not treat the reported F1 score as a protocol-controlled estimate.

**5. Reference [34]** (Christoffersen & Gonçalves, *Estimation risk in financial risk management*) is cited as the source for PRD and RMSE. It does not define PRD.

**6. Reference [35]** (Mittnik & Rachev, *Alternative multivariate stable distributions*) is cited as the source for the Fréchet distance. It is not a source for that metric.

**7. References [30] and [31] are misattributed in the text.** Reference [30] is described in the body as "WaveGAN"; the reference-list entry is Binkowski, Marti & Donnat, *Autoregressive convolutional neural networks for asynchronous time series* (ICML 2018). Reference [31] is described as "SampleRNN"; the entry is Schreyer, Sattarov, Reimer & Borth, *Adversarial learning of deepfakes in accounting* (2019).

**8. Reference [33]**, cited for the Pearson correlation formula, is a commercial tutoring website and is mislabelled in the reference list.

**9. Code availability.** Sec. V-3 states that code would be made publicly available under the GitHub account `WilfredoTovar`. A repository was published and subsequently withdrawn by the author. Recovery is being attempted; any recovered code will be released with the follow-up study.

---

**On the scope of these corrections.** Items 1–4 concern the reporting of the study rather than its architecture; the model described in Secs. V and VI is as published. Items 5–8 are citation errors. None of the corrections is intended to defend the original evaluation methodology, which the author now regards as inadequate for the claims made — in particular the use of PRD, pointwise RMSE and the curve Fréchet distance to assess a generator of stochastic paths, and the reporting of directional F1 without base-rate comparison, significance testing, transaction costs, or naive baselines. That assessment, and a re-evaluation of the study under a pre-registered protocol, is the subject of separate work in preparation.

**Citing this paper.** Authors citing this work are asked to cite v3 and to note items 1 and 2 where market coverage or sample size is material to the citation.

# Deep Learning Based on Generative Adversarial and Convolutional Neural Networks for Financial Time Series Predictions

Wilfredo Tovar Hidalgo
School of Information Technology
Carleton University
Ottawa, Canada
wilfredotovarhidalgo@cmail.carleton.ca

## ABSTRACT

In the big data era, deep learning and intelligent data mining technique solutions have been applied by researchers in various areas. Forecast and analysis of stock market data have represented an essential role in today's economy, and a significant challenge to the specialist since the market's tendencies are immensely complex, chaotic and are developed within a highly dynamic environment. There are numerous researches from multiple areas intending to take on that challenge, and Machine Learning approaches have been the focus of many of them.

There are multiple models of Machine Learning algorithms been able to obtain competent outcomes doing that class of foresight. This paper proposes the implementation of a generative adversarial network (GAN), which is composed by a bi-directional Long short-term memory (LSTM) and convolutional neural network(CNN) referred as Bi-LSTM-CNN to generate synthetic data that agree with existing real financial data so the features of stocks with positive or negative trends can be retained to predict future trends of a stock. The novelty of this proposed solution that distinct from previous solutions is that this paper introduced the concept of a hybrid system (Bi-LSTM-CNN) rather than a sole LSTM model. It was collected data from multiple stock markets such as TSX, SHCOMP, KOSPI 200 and the S&P 500, proposing an adaptative-hybrid system for trends prediction on stock market prices, and carried a comprehensive evaluation on several commonly utilized machine learning prototypes, and it is concluded that the proposed solution approach outperforms preceding models. Additionally, during the research stage from preceding works, gaps were found between investors and researchers who dedicated to the technical domain.

## CCS CONCEPTS

• Big Data • Deep Learning • Data Mining • Long short-term memory (LSTM) networks •Generative Adversarial Network (GAN) • Convolutional Neural Network (CNN).

## KEYWORDS

Deep Learning, Data Mining, Long short-term memory (LSTM) networks, Generative Adversarial Network (GAN), Convolutional Neural Network (CNN), Bi-LSTM-CNN , Forecast and analysis of the stock market.

## I. INTRODUCTION

Prognostications on financial stock markets have been an object of studies, but given its innate complexity, dynamism and chaoticness, it has proven to be a particularly challenging task. The abundance of variables and sources of information held is immense. For decades, there have been studies in Science regarding the possibility of such a feat, and it is notable in the related literature that most prediction models neglect to present accurate predictions in a broad sense. Nevertheless, there is a tremendous number of studies from various disciplines seeking to take on that challenge, presenting a large variety of approaches to reach that goal.

Since the forecast of the financial stock market has drawn attention from industry to academia as in [1] [2]. Various machine learning algorithms, such as genetic algorithms, support vector machines, neural networks, among others, have been in use to predict the market variances. Recurrent neural networks (RNNs) hold a robust model for treating sequential data such as sound, time-series data or written natural language. Some designs of RNNs were used to predict the stock market as in [3] [4].

The traditional approach is to use Machine Learning algorithms to learn from historical price data to predict projected values to expedite automated analysis, such as support vector machines[5], decision trees[6] LSTM based method [7] and recently developed deep learning methods[8]. However, most of these approaches demand extensive amounts of labelled data to train the model, which is a practical intricacy that still needs to be resolved. For instance, large volumes of labelled Stocks Operations data are typically required as training representations for financial tendencies distribution systems. Furthermore, when machine learning approaches are employed to specific strategies of investment operations, such as scalping, day trading, swing trading, or position trading, the collected training data are often categorized based on the specific features of the stocks, bonds, currencies or others products, such as their price, return on equity, PEG ratio, earnings per share, book value, among others. Thus, the problems caused by lacking useful Stock data are exacerbated before any subsequent analysis. Additionally, maintaining the secrecy of financial information is always an issue that cannot be ignored. In consequence, an optimal solution is to generate synthetic data and combined it with publicly available data to satisfy the requirements for analysis and predictions.

This proposal intends to maintain that direction but studying a specific hybrid method employing a GAN architecture, which is compounded by a generator and a discriminator. In the generator component, the inputs are static data values obtained from a Gaussian distribution. We produce two layers of bidirectional long short-term memory (Bi-LSTM) networks, which possesses the advantage of selectively preserving the past and current information. Furthermore, to prevent over-fitting, a dropout layer was built. In the discriminator component, we classify the extracted data by employing a design based on a convolutional neural network (CNN). The discriminator incorporates two sets of convolution-pooling layers combined with a wholly connected layer, a softmax and an output layer from which a bi-nary rate is defined grounded on the estimated one-hot vector. In this project, were used multiple cleaned datasets organized and formed independently. However, the data is an open-sourced from multiple markets obtained at the Yahoo Finances platform; this paper illustrates the data collection details in section IV.

Our model outperformed others similar traditional two methods, the deep recurrent neural network-autoencoder (RNN-AE)[9], the RNN-variational autoencoder (RNN-VAE)[10], and the sole LSTM approach. Additionally, during the research stage from preceding works, gaps were found between investors and researchers dedicated to the technical domain.

## II. PROBLEM STATEMENT

This research project determines four main research questions fundamental on an ample preceding literature review and suggests a comprehensive solution supported by a full evaluation that intends to solve the following research interrogations.

1. What is the best algorithm for prediction short term price trend?

Assume from previous works, researchers are pursuing the exact price prediction. While this will be intended to crumble the problem into predicting the trend and then an exact number, this paper will concentrate on the first step. Therefore, this objective will be converted to solve a binary classification problem, meanwhile discovering an effective way to eliminate the negative effect brought by the high level of statistical noise. This approach will decompose the complex problem into sub-problems that produce fewer dependencies and solve them one by one, then compile the resolutions into an ensemble model as an aiding system for investing behavior reference.

2. Is a hybrid machine learning system required for multifactorial times series prediction?

In deep learning, patterns are recognized as black-box with billions of parameters. However, customized engineering systems are still vital to enhance design precision. For instance, neural network like Long-Short Term Memory of deep learning can connect contemporary and preceding events while approaches such Structural Time Series Designs demonstrate a high level of accuracy and only depends on the previous event. Therefore, combining different approaches and techniques is fundamental to address the limitations and enhance accuracy.

3. How machine learning innovations benefit the model's forecast precision in the Financial Domain?

From the plentiful previous publications evaluate, we can assume that stock price data are embedded with a high level of statistical dissonance, and there are relationships among features, which makes the price prediction particularly complex to predict. That is the principal cause for most of the previous relate works introduced the feature engineering part as an optimization module.

4. How do innovations from economic-financial field privileges prediction model design?

Distinct from previous works, besides the standard evaluation of data models such as the training costs and rates, this paper pretends to evaluate the effectiveness of recently added features that can be extracted from the financial domain. It will include some features from the financial domain. While is obtained some specific findings from previous research works, and the related raw data needs to be processed into usable features. After extracted related features from the financial domain. This paper will combine the features with other standard technical indices for voting the features with high impact.

There are numerous features said to be valid from the financial area. Thus, how to accurately transform the conclusions from the financial field to a data processing module of the proposed system design is a covered research question that must be facing within this research.

## III. SURVEY OF RELATED WORK

Due to its complexity and dynamism, there has been a constant dispute on the predictive performance of several stock returns predictors. In regard to computational intelligence, there are several studies assessing various methods in order to accomplish accurate predictions on the stock market. K. Kim in [11] Uses statistical learning by using algorithms like Support Vector Machines (SVM), Estrada [12] Proposed a new approach based on collective intelligence, including neural networks, component modelling, textual analysis based on news data.

There are effective works related to deep learning in stock markets; there are some examples like Sharang Et al. [13] published a study where is performed on the application of a Deep Belief Network (DBN), which is composed of stacked Restricted Boltzmann Machines, coupled to a Multi-level Perceptron (MLP) and using long-range log returns from stock prices to predict above-median returns. Heaton Et al. [14] implemented DBN, using price history in addition to technical indicators as input, in a similar approach to this project. Both of those works present improved results matched to their baselines, as well as in Greff Et al. [15] presented a survey in deep learning methods applied to finance is done and their improvements discussed.

Lei in [16] utilized Wavelet Neural Network (WNN) to prognosticate stock price trends, this approach also included Rough Set (RS) for attribute reduction as an optimization, it was employed to reduce the stock price trend characteristic dimensions and to define the composition of Wavelet Neural Network. The prototype evaluation was proved on different stock market indices, and the result was satisfying with generalization. It demonstrates that Rough Set would effectively decrease the computational complexity. However, the author only emphasized the parameter adaptation in the discussion part while it did not define the vulnerability of the model itself. In consequence, during the literature review, it was found that the evaluations were conducted on indices, and performance varies if it is implemented on specific stocks. The features table and calculation formula are worth taking as a reference.

Gheyas and Smith [17] states that the shortcomings of an individual machine learning system which straight influence the forecasting precision. It is due that these designs invariably undergo from the parametric structure obstacles such as choosing multiple parameters, local optima, and overfitting. Evans et al. [18] suggest that forecasting performance for time series data depends on factors such as the nature of data and the selected methodology. It is essential to adjust the market behaviour based on the most suitable configuration to be chosen for each case.

To overcome this shortcoming, various researchers suggested the application of hybrid approaches. Nayank et al. [19] proposed a hybrid tool for stock market averages forecast. This approach combines SVM with K-nearest neighbor (KNN) approach to determine the closing value and identify the trend, the volatilization, and the impulse of the stock market.

Cavalcante and Oliveira [20] produced an original trading prototype, including the application of ELM and OS-ELM ensembles, to negotiate in the stock market autonomously. Ballings et al. [21] presented a related research on the performance of Ensemble systems over individual classifier designs in forecasting stock value tendencies. Ensemble systems such as the AdaBoost, kernel factory, and random forest beat the individual classifier as ANN, SVM, KNN, and logistic regression.

McNally et al. in [22] Presented a work based on Recurrent neural network (RNN) and LSTM to prognosticating cryptocurrencies by applying the Boruta algorithm for the feature engineering part, which operates comparable to the random forest classifier. They also employed Bayesian optimization to select LSTM parameters. The principal dilemma of their approach is overfitting. The research problem of predicting the cryptocurrencies price course has remarkable relations with stock prediction. There are unknown characteristics and noises inserted in the price data, and treat the research problem as a time sequence problem.

Fischer and Krauss in [23] applied long short-term memory (LSTM) successfully on financial stock market prediction. Their dataset was compounded by the S&P 500 index, where the lists were consolidated into a binary matrix to eliminate survivor bias. The authors also applied effectively an optimizer called "RMSprop, "which is a mini-batch variant of rprop. This paper offers a robust notion of times series predictions, but it is not suitable for PRICE Trend Forecast. The principal strength of this research is that the designers adopted the most advanced deep learning technique by 2018 to offer financial market prognostications. However, they relied on the LSTM system, causing a misconception of its practical implementation, produced by a lack of background and understanding of the financial domain. Though the LSTM outperformed the conventional DNN and logistic regression algorithms, the author did not consider the effort and application to train an LSTM with long time dependencies. Consequently, Their approach showed that LSTM is suitable for financial time series prediction tasks that are a different domain than Short-term PRICE Trend Forecast, which is one of the aim of the proposed approach on this paper.

### *A. Gap Analysis*

This section illustrates the gaps found from the information and comparison of contents of related researches.

The gaps found among the research articles in the finance field and technical domain are that Technical-Scientific research papers tend to focus on building the models more effectively, and occasionally, they do not consider the applicability of their designed system and models or disregards the implementation of hybrid machine learning techniques to optimize those systems. In research papers, there is a process of selecting the features that will be covered and are examined and mentioned from preceding works and go through the feature selection algorithms. While in the financial domain, the researchers show more interest in behaviour analysis, since they may affect the stock performance. During their studies, they frequently conduct a full statistical analysis based on a particular dataset and decide new features rather than performing feature choosing.

From those findings, It may be inferred that the financial domain are rarely being deep investigated in the technical domain, and the financial domain research paper also hardly introduced simples or hybrid machine learning or deep learning algorithms to train their model.

### *B. Trend Models of Structural Time Series Models*

***Generative-Adversarial Networks.*** The GAN are deep generative systems that diverges from other generative designs such as autoencoder in terms of the techniques applied for producing data and is essentially composed of a generator and a discriminator. The generator generates data based on sample data points that reflect a Gaussian pattern and learn from the feedback provided by the discriminator. The discriminator receives the odds spread of the authentic data and produces a true-or-false condition to resolve whether the created data are genuine. The two sub-models holding the generator and discriminator give a merging state by performing a zero-sum game. Figure 1 shows the formula of GAN. The result achieved by GAN can be seen as a min-max optimization method. The objective function is:

$$\min_G \max_D V(D, G) = E_{x\sim p_{data}(x)}[\log D(x)] + E_{z\sim p_z(z)}[\log(1 - D(G(z)))],$$

Figure 1.

Assuming that D is the discriminator and G is the generator. At the time the distribution of the real data is equal to the combination of the produced data, the output of the discriminator can be viewed as the optimal value. Generative-Adversarial Networks has been successfully applied in several areas such as natural language processing[24][25].

**RNN.** The recurrent neural network has been extensively applied to work on tasks of processing time-series data[26]. RNN typically involves an input layer, a deep/hidden layer, and an output layer, where the deep/hidden event at a particular time *t* is defined by the input at the current time as well as by the deep/hidden state at a previous time:

$$h_t = f(W_{ih}x_t + W_{hh}h_{t-1} + b_h),$$

Figure 2.

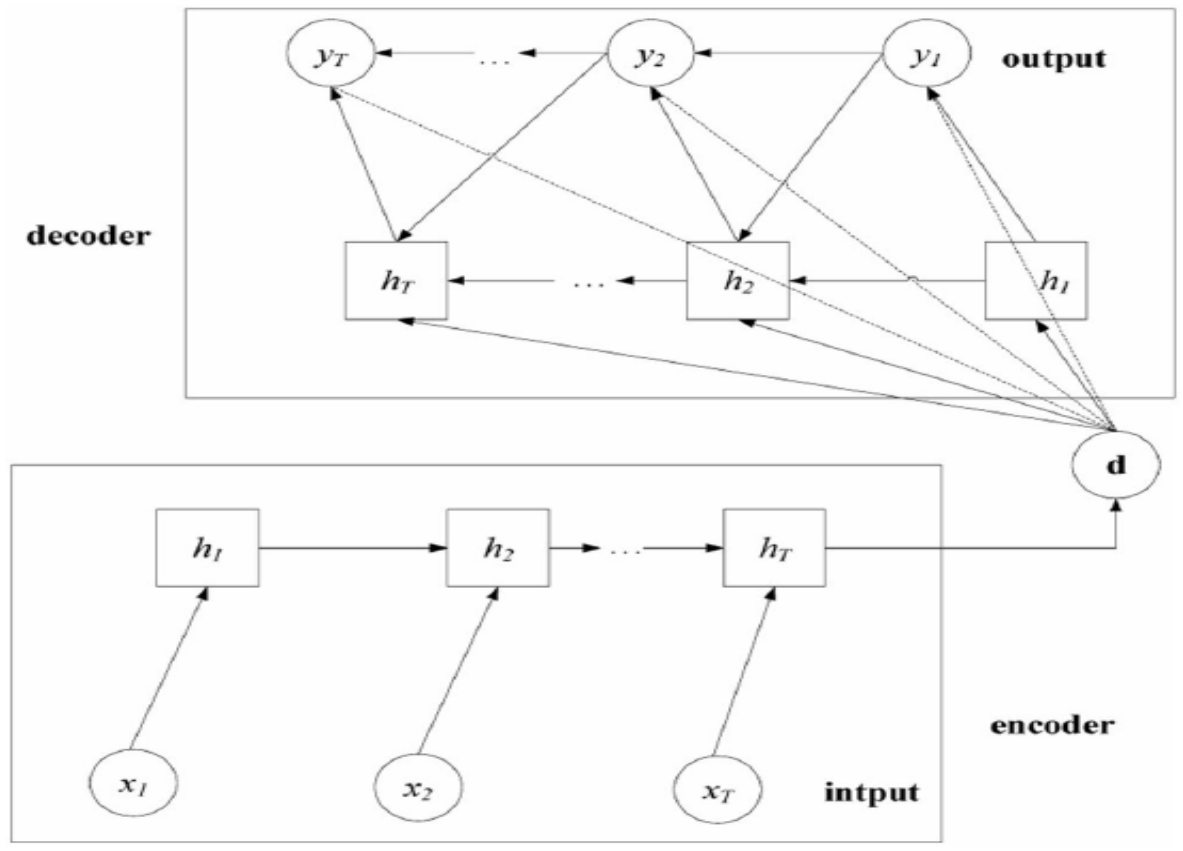

Figure 3. RNN-AE Design.

$$o_t = g(W_{ho}h_t + b_o),$$

Assuming that *f* and *g* are the starting/activation functions, xt and ot are the input and output at time *t*, individually, *ht* is the hidden state at time *t*, W{*ih*,*hh*,*ho*} describe the weight patterns that join the input layer, hidden layer, and output layer, and b{*h*,*o*} indicate the base of the hidden layer and output layer. RNN is exceptionally suitable for short-term minor problems; however, it is incompetent with long-term dependent problems. The long short-term memory (LSTM) and gated recurrent unit (GRU) were developed to master the shortcomings of the Recurrent Neural Networls, including gradient expansion or gradient disappearance throughout the training process. The LSTM is a modification of an RNN. It is suitable for processing and prognosticating major events with long pauses and delays in time series data by using other architecture denominated the memory cell to store earlier obtained information. The GRU is also a modification of an RNN, which links the neglect gate and input gate into an update gate to manage the volume of information analyzed from previous time flows at the current time. The reset gate of the GRU is employed to master how much data from previous times is ignored.

**RNN-Autoencoder and RNN-Varational Autoencoder.** Those are generative models introduced before GAN. Additionally, used for creating data[27], they were employed to dimensionality reduction[28]. RNN-AE is an extension of the autoencoder model where both the encoder and decoder operate with RNNs. The encoder outputs a deep, latent code **d**, which is one of the decoder's input values. In opposition to the encoder, the decoder's output and hidden/deep state at the current time are depending on the output at the prevailing time and the hidden/deep state of the decoder at the preceding time as well as on the latent code d. The purpose of RNN-AE is to secure the decoder's raw data and output as similar as reasonable. It can be seen on Figure 3 that shows the RNN-AE framework.

VAE is a modification of autoencoder where the decoder no longer outputs a hidden/deep vector, but alternately returns two vectors containing the median vector and variance vector. A facility named the re-parameterization trick is practiced to re parameterize the random code **z** as a deterministic code, and the hidden/deep latent code **d** is obtained by linking the median vector and variance vector:

$$\mathbf{d} = \mu + \sigma \odot \varepsilon,$$

Where μ is the median vector, σ is the variance vector, and ε ~ N (0, 1).

RNN-VAE is a modification of VAE where an independent/single-layer RNN is applied in both the encoder and decoder. This design is proper for discrete tasks such as progression-to-progression or sequential learning and decision generation.

**Formation of Time Series Data.** As far as the literature were reviewed**,** there is a single published study implementing this relevant techniques of deep learning to produce or synthesize financial data [29], where the authors determine that GANs can learn and accurately reproduce intricate features distribution and relationships between features of real modelling data[29]; however, their model fails on training synthetic data and "showed slightly worse performances on out-of-sample validation data[29]", In consequence, this paper intend to overcome the performance of their model with the proposed hybrid solution.

Techniques for creating raw financial data in waveforms were principally based on the practice of autoregressive designs, such as [30] and SampleRNN[31], both of them utilizing conditional probability models, which means that at time t each sample is produced according to all samples at former time impressions. However, autoregressive frames have the tendency to result in slow creation since the output financial representations have to be supply back into the model once each time, while GAN is able to bypass this limitation by continually performing an adversarial training to obtain the distribution of produced results and real data as close as possible.

In this study, each point examined from the market is expressed by a single-dimensional vector of the time-step and leads. WaveGAN[30] uses a single-dimensional filter of length and a large up-sampling factor. However, it is imperative that these two procedures have an equal number of hyper parameters and mathematical calculations. According to the foregoing analysis, our design of GAN will embrace deep LSTM layers and CNNs to optimize formation of time series sequence.

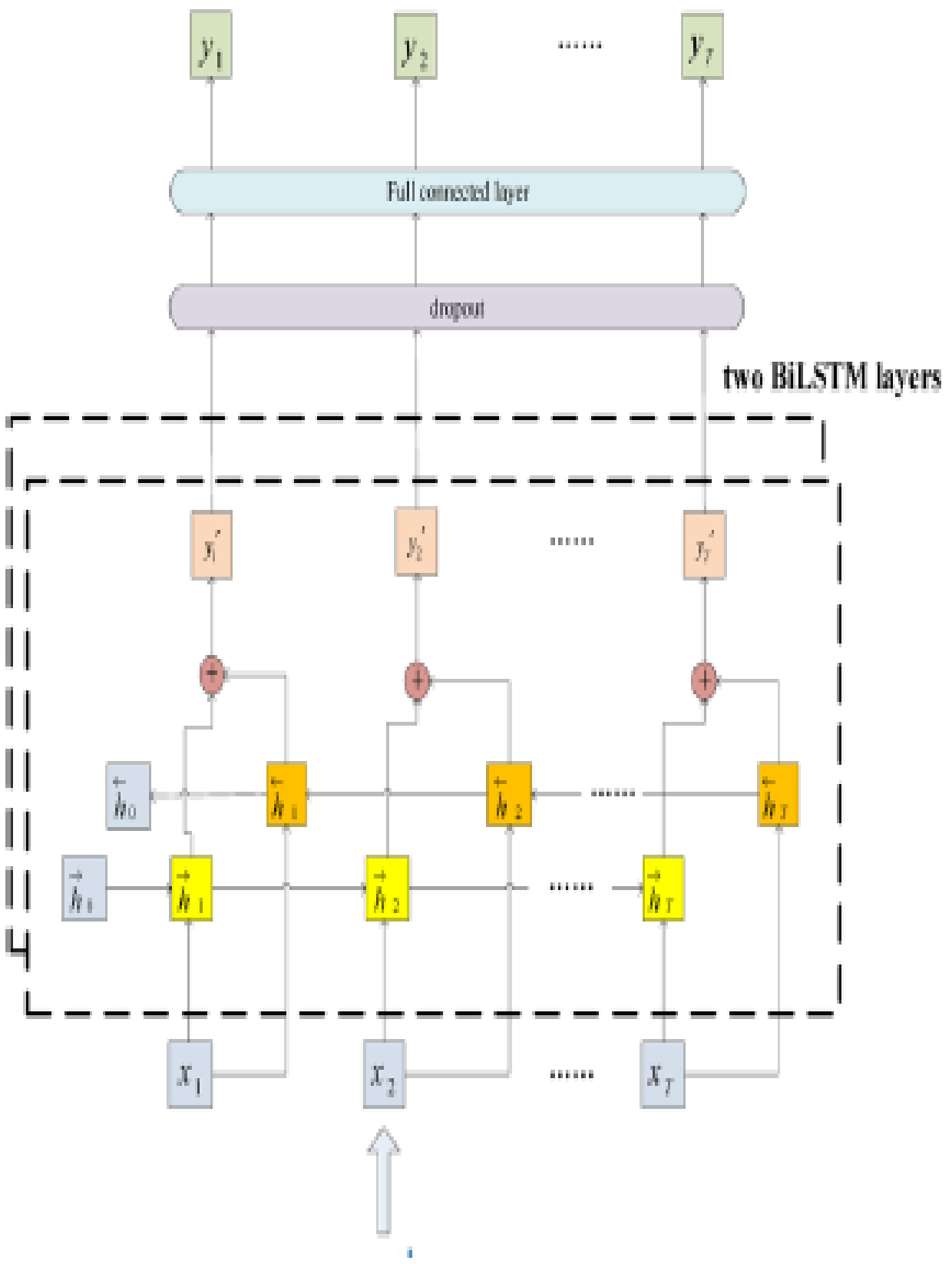

Figure 4. Illustration of Generator's Design.

# IV. PROCEDURE OF THE STUDY

## 1. Data

*A. Datasets Preparation and Description*

Stock price and volume are the two criteria that indicate the stock's market performance in the present, past and upcoming future. In consequence, employing a technical analysis with the data may be employed to train the design and generate an intelligent machine for stock prediction.

The experiments used historical closing price data from the following stock markets indices, i.e., TSE, SHCOMP, and the S&P 500, and included all stocks in each index from 31 January 2012 to 31 July 2018 on a working day basis. These data were obtained from the Yahoo financial website (https://finance.yahoo.com/). This dataset consists of 2954 stocks. It was collected everyday price data, stock ID, reopening history, and the top 5 main shareholders. The data used were from 31 January 2012 to 31 July 2018, since exercising with more up-to-date data would help the study result.

*B. Dataset Structure Design.*

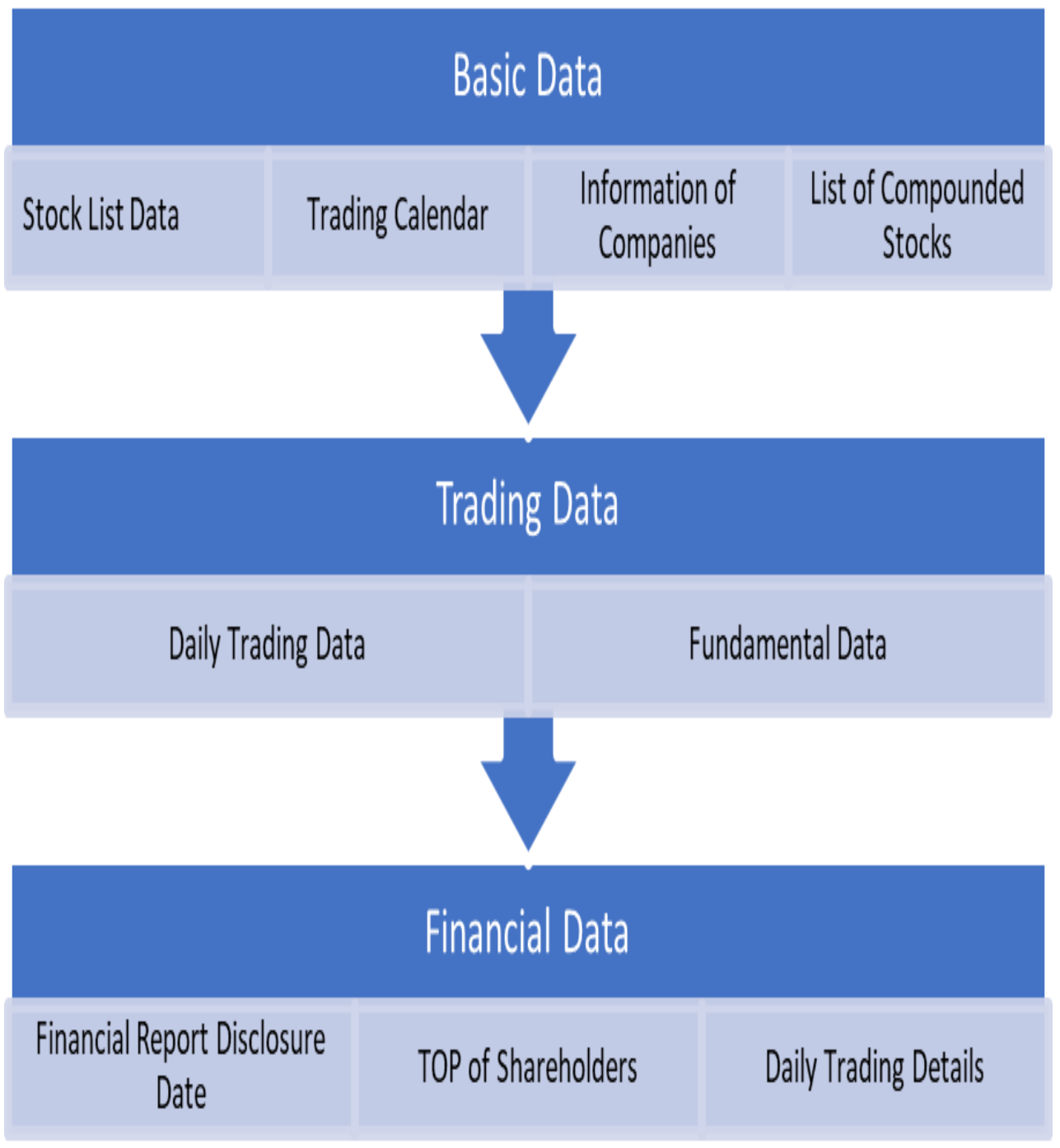

*Figure 5. Dataset Structure Design.*

Figure 5 shows the total of the data on the dataset tables. There are three classes of data in this dataset: basic data, trading data, finance data.

The totality of the data table can be associated with a track denominated "S-ID. Or Stock-ID" It is an individual stock identifier recorded in the commodities markets. Basic Data is the essential information that the users or researchers might require during the exploration process of the data. It is compounded for Stock-List

Data, Trading Schedules, Primary Information of Listed Corporations, etc.

Trading data is the value and trade-related data for a financial instrument published by a trading regulator or market. Finance data, the information considered on this case was based on the income statement and balance sheet of every Stock-ID.

### *C. Technical Background & Indices [32]*

In this section, the most regularly employed technical indices are illustrated.

1) Stochastic indicator K.

The n-day stochastic indicator K is represented as:

$$K_n_i = \frac{2}{3} \leftrightarrow K_n_{i-1} + \frac{1}{3} \leftrightarrow \frac{CP_i - LP_n_i}{HP_n_i - LP_n_i} \leftrightarrow 100$$

2) Stochastic indicator D.

The n-day stochastic indicator D

$$D_n_i = \frac{2}{3} \leftrightarrow D_n_{(i-1)} + \frac{1}{3} \leftrightarrow K_n_i$$

Where K_ni is the n-day stochastic indicator K of day i.

3) Williams overbought/oversold index

The n-day Williams overbought/oversold index is a momentum indicator that measures overbought and oversold levels.

$$WMS\%R_n_i = \frac{HP_n_i - CP_i}{HP_n_i - LP_n_i}$$

4) Commodity channel index

The commodity channel index is used to identify cyclical turns in commodities.

It is define the typical price as the formula below:

$$TP_i = \frac{HP_i + LP_i + CP_i}{3}$$

Then it is calculate the n-day simple moving average of the typical price:

$$SMATP_n_i = \frac{\Sigma_{j=i-n-1}^{i} TP_j}{n}$$

And the n-day mean deviation is noted by MD_n:

$$MD_n_i = \frac{\Sigma_{j=i-n-1}^{i} |TP_j - SMATP_n_i|}{n}$$

The n-day commodity channel index is calculated as:

$$CCI_n_i = \frac{TP_i - SMATP_n_i}{0.015 \leftrightarrow MD_n_i}$$

5) Relative strength index, is a momentum indicator employed in technical analysis that estimates the extent of recent price fluctuations to estimate overbought or oversold in the price of a stock..

$$G_i = \begin{cases} CP_i - CP_{i-1} & \text{if } CP_i > CP_{i-1} \\ 0 & \text{otherwise} \end{cases}$$

6) Moving average convergence/divergence

It is a trend-following momentum pointer that determines the correlation between two moving proportions of a security's price.

First step is to calculate the demand index (DI):

$$DI_i = (HP_i + LP_i + 2 \leftrightarrow CP_i)/4$$

We also need to calculate the 12-day and 26-day exponential moving average:

$$EMA_12_i = \frac{11}{13} \leftrightarrow EMA_12_{i-1} + \frac{2}{13} \leftrightarrow DI_i$$

And

$$EMA_26_i = \frac{25}{27} \leftrightarrow EMA_26_{i-1} + \frac{2}{27} \leftrightarrow DI_i$$

Hence, we use DIFi to indicate the difference between EMA_12 and EMA_26:

$$DIF_i = EMA_12_i - EMA_26_i$$

The MACDi is calculated as below:

$$MACD_i = \frac{8}{10} \leftrightarrow MACD_{i-1} + \frac{2}{10} \leftrightarrow DIF_i$$

7) 10-day moving average

It is the **average** based on the closing value of a stocks over the last **10** periods:

$$MA_10_i = \frac{\Sigma_{j=i-9}^{i} CP_j}{10}$$

8)Momentum

Momentum measures change in stock price over last n days.

$$MTM_i = \frac{CP_i}{CP_{i-n}} \times 100$$

9) Rate of Change

The n-day rate of change measures the percent changes of the current price relative to the price of n days ago and is calculated by:

$$ROC_n_i = \frac{CP_i - CP_{i-n}}{CP_{i-n}} \times 100$$

10) Psychological line

The psychological line is a volatility indicator based on the number of time intervals that the market was up during the preceding period and is calculated by:

$$PSY_n_i = \frac{TDU_n_i}{n} \times 100\%$$

The TDU_ni is the total number of days that has price rises in previous n days.

11) AR

n-day A ratio means the n-day buying/selling momentum indicator which is calculated as:

$$AR_n_i = \frac{\Sigma_{j=i-n-1}^{i}(HP_j - OP_j)}{\Sigma_{j=i-n-1}^{i}(OP_j - LP_j)}$$

12) BR

n-day B ratio means the n-day buying/selling willingness indicator and is defined as:

$$BR_n_i = \frac{\Sigma_{j=i-n-1}^{i}(HP_j - CP_{j-1})}{\Sigma_{j=i-n-1}^{i}(CP_{j-1} - LP_j)}$$

13) Volume ratio

The n-day volume ratio is calculated by:

$$VR_n_i = \frac{TVU_n_i - TVF_n_i/2}{TVD_n_i - TVF_n_i/2} \times 100\%$$

Where the TVU represents the total trade volumes of stock price rising, and TVD is the total trade volumes of stock prices falling, TVF represents the total trade volumes of stock prices holding in previous n days.

14) Accumulation/distribution oscillator

$$AD_i = \frac{HP_i - CP_{i-1}}{HP_i - LP_i}$$

15) 5-day bias

The 5-day bias is the deviation between the closing price and the 5-day moving average MA_5

$$BIAS_5_i = \frac{CP_i - MA_5_i}{MA_5_i}$$

*D. Data Representation.*

In this section, the raw data are converted to extract the appropriate properties for the examination to be performed in a more precise way. It was calculated the important characteristics for analyzing the data:

• Interval open price (IOP)

• Interval close price (ICP)

• Interval low (IL) is the lowest price in a interval of trading.

• Interval high (IH) is the highest price in a interval of trading.

• Interval average (IA) is the average price of each trading interval.

• Volume is the stock volume at the end of each trading interval.

• Interval price (IP) is the stock price at the end of each trading interval.

• Time is the current time at the which trading of each interval closes.

• Date is the current date at which the trading of each interval closes.

*E. Starting point window approach*

Window dimension for data processing is one of the vital determinants to be fixed to obtain the precision in a forecasting model. Assume the window dimension of one session is 40 interim with a time interim of 10min for each interim is to be processed N then 40 interval data might not be much adequate for feature production in the intra-day forecast since it is said to be an outdated data. If the window dimension of one period is obtained as 5 intervals, it does not examine the preceding trend and market fluctuation, which influences the precision of the model. In this research work, the window dimension of one period has 14 intervals with a time determination of 10min as shown in Figure 3(a)is illustrated at figure 6.

It is proposed as an ideal stage to do the technical analysis with the features measured as the technical indicators which inform the market trends and gives the precise course of the price variation. After each repetition, the sliding window data is refreshed, including the subsequent interval data being added to the window and the oldest interim data is staying discarded from the window. Consequently, it can be stated that it is dynamically renewing the data with the supplied window size of 14 intervals as one period for feature creation that includes information of 140minmarket data fluctuation.

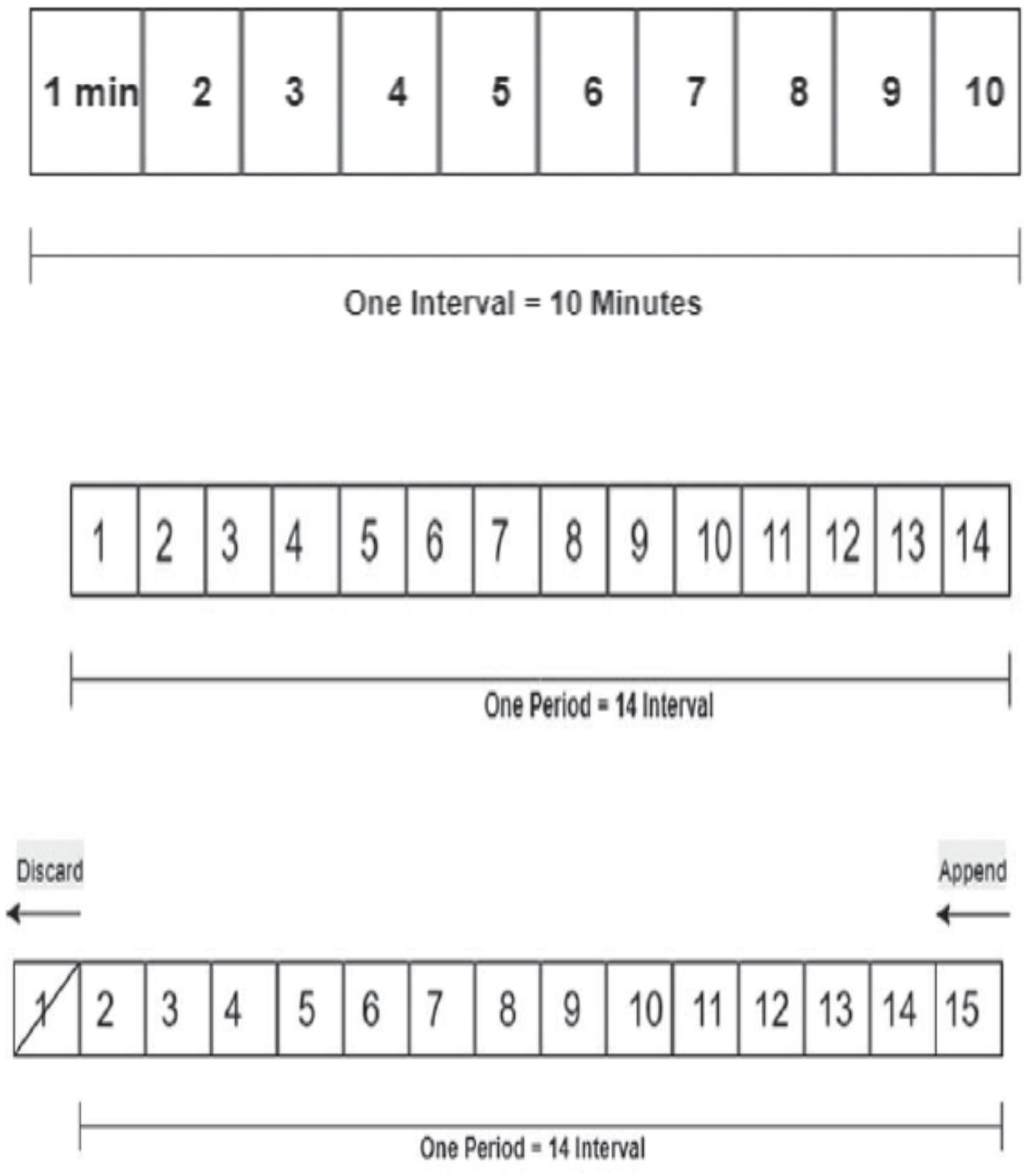

Figure 6. Time Resolution of Intervals

## V. PROPOSED SOLUTION

This research presents a GAN based design for generating time series prediction. The design contains a generator and a discriminator. The input to the generator includes a series of sequences where each sequence is made of 3000 noise points. The output is a generated financial sequence with a length that is also set to 3000. The input to the discriminator is the produced result and the real Financial data, and the output is D(x) ? {0, 1}. In the training process, G is originally fixed, and we train D to increase the possibility of selecting the correct label to both the realistic points and generated points. We then train G to reduce log(1 - D(G(z))). The objective function is described as:

$$\min_{G_\theta}\max_{D_\phi} L_{\theta;\phi} = \frac{1}{N}\sum_{l=1}^{N}[\log D_\phi(x_l) + (\log(1 - D_\phi(G_\theta(z_l))))],$$

Where N is the representation of points in this case 3000 points for each distribution/sequence in this study, and θ and φ denote the assortment of parameters. As CNN does not have recurrent links like ignoring units as in GRU or LSTM, the prototypes' training method with a CNN-based discriminator is often faster, particularly in the event of long series of data modelling.

**Generator's Design**.

A set of data points with high sonance or volatility that reflect a Gaussian dispersion are filled into the generator as a fixed interval sequence. It is assume that each noise point can be expressed as a d-dimensional one-hot vector and the length of the sequence is T. In consequence, the size of the input matrix is T × d. The generator contains two BiLSTM layers, each having 90 cells. A dropout layer is merged with a fully connected layer. The illustration of the generators' architecture is shown in Fig. 4.

The current deep/hidden positions are dependents on two deep/hidden positions, one from progressive LSTM and the other from regressive LSTM. Eqs 1 and 2 are employed to determine the deep/hidden positions from a pair of equal directions and Eq. 4 determines the output of the first Bi-LSTM layer at time t:

$$\overrightarrow{h}_t^1 = \tanh\left(W^1_{i\overrightarrow{h}} x_t + W^1_{\overrightarrow{h}\overrightarrow{h}} \overrightarrow{h}^1_{t-1} + b^1_{\overrightarrow{h}}\right), \quad (1)$$

$$\overleftarrow{h}_t^1 = \tanh\left(W^1_{i\overleftarrow{h}} x_t + W^1_{\overleftarrow{h}\overleftarrow{h}} \overleftarrow{h}^1_{t+1} + b^1_{\overleftarrow{h}}\right), \quad (2)$$

$$y_t^1 = \tanh\left(W^1_{\overrightarrow{h}o} \overrightarrow{h}_t^1 + W^1_{\overleftarrow{h}o} \overleftarrow{h}_t^1 + b_o^1\right), \quad (3)$$

Where the output depends on →h t and ←h t, and h0 is initialized as a zero vector. Likewise, it is obtained the out-put at time t from the second Bi-LSTM lay

$$y_t = \tanh\left(W^2_{\overrightarrow{h}o}\ \overrightarrow{h^2_t} + W^2_{\overleftarrow{h}o}\ \overleftarrow{h^2_t} + b^2_o\right). \quad (4)$$

To block slow gradient reduction due to parameter expansion in the generator, it was added a dropout layer and set the odds to 0.4. The output layer is a bi-dimensional vector where the primary component denotes the time step, and the second factor indicates the lead.

**Discriminator's Design**

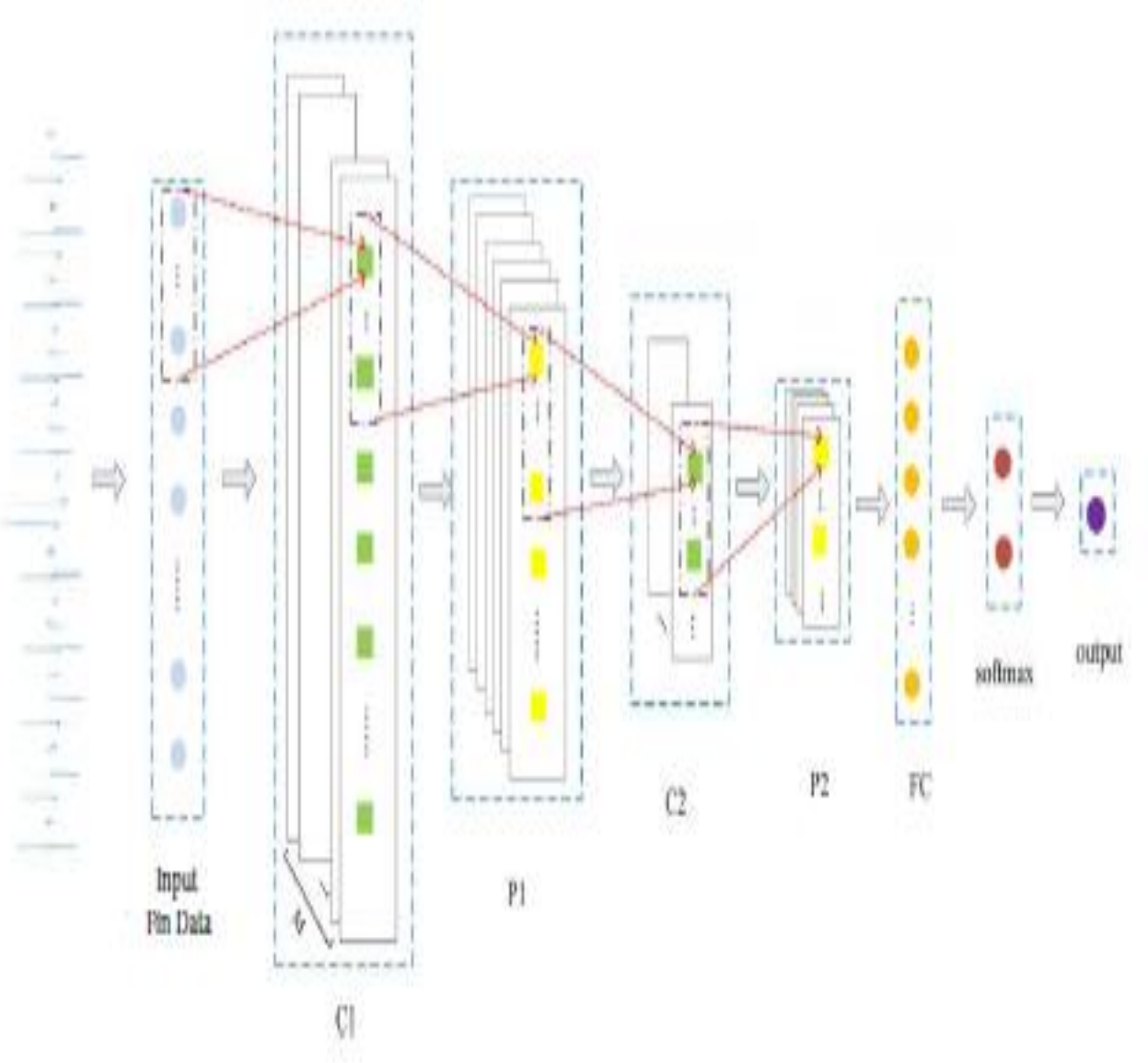

Figure 7. Illustration of Discriminator's Design.

In Figure 7. It is showed the structure design of the Discriminator. The set of red dashed tracings on the left indicates a mapping class, symbolizing the place where a filter is transferred—the ones on the right exhibit the value obtained by employing the convolutional process or the pooling procedure. The series containing Values data points can be viewed as a time series sequence (a typical graph model demands both a vertical and a horizontal convolution) rather than an image, so only one-dimensional (1-D) convolution requires to be included. It is expected that an input sequence x1, x2, … xT comprises T points, where a d-dimensional vector outlines each. Additionally, it was established the dimension of filter to h*1, the dimension of the stride to k*1 (k ≪ h), and the representation of the filters to M. Accordingly, the output size from the primary convolutional-layer is M * [(T − h)/k + 1] * 1. The window for the filter is:

$$x_{l:r} = x_l \oplus x_{l+1} \oplus x_{l+2} \oplus \ldots \oplus x_r.$$

The conditions of *l* and *r* are determined by:

$l = k * i + 1 - k$ $l \in [1, T - h + 1]$,

$r = k * i - k + h$ $r \in [h, T]$,

where $1 \leq k^*i+1 \leq T-h+1$ and $h \leq k^*i-k+h \leq T$ $(i \in [1, (T-h)/k+1])$.

The returned convolutional sequence $c = [c_1, c_2, \ldots c_i, \ldots]$ with each $c_i$ is calculated as

$$c_i = f(w * x_{l:r} + b),$$

$$c_i = f(w * x_{l:r} + b),$$

where $w \in R_{h \times d}$ a combined weight matrix, and *f* represents a nonlinear activation function. The replacement layer is the max pooling layer with a window size of *a*1* and stride size of *b*1*. Each output from pooling $p_j$ for the rebounded pooling determines the sequence $p = [p_1, p_2, \ldots p_j \ldots]$ that is:

$$p_j = \max(c_{bj+1-b}, c_{bj+2-b}, \ldots c_{bj+a-b}).$$

After performing double sets of processes for convolution and pooling, it is attached completely connected to a layer that joins it to a softmax layer, where the output is a one-hot vector. The two components in the vector represent the probability that the input is true or false. The function of the softmax layer is:

$$\sigma(z)_j = \frac{e^{z_j}}{\sum_{k=1}^{2} e^{z_k}} (j = 1,\ 2).$$

In Figure 8, C1 layer is a convolutional layer; the dimension of each filter is represented by 120*1, the quantity of filters is 10, and the size of stride is 5*1. The output size of C1 is measured by:

$$\frac{(W, H) - F + 2P}{S} + 1,$$

Where (W, H) denotes the input volume size (1*3120*1), F and S indicate the dimension of kernel filters and length of stride each, and P is the number of zero paddings, and it is compounded set to 0. Consequently, the output size of C1 is 10*601*1. Figure 7, the P1 layer is a pooling layer where the size of each window is 46*1 and the size of stride is 3*1. The output size of P1 is calculated by:

$$\frac{(W, H) - F}{S} + 1,$$

Where (W, H) describes the input dimensions size (10x601x1), F and S denote each window's size and length of stride. Therefore, the determined output size of P.1 is (10*186*1). The principle of compu-parameters of convolutional layer are C2 and pooling layer P2 is the equal as that of the preceding layers. It requires to be highlighted that the amount of kernels filters of C2 is set to 5. With pairs of convolution-pooling processes, we obtained the output size as 5*10*1. An utterly connected layer, which includes 25 neurons, joins with P2. The ultimate layer is the softmax-output layer, which outputs the decision of the discriminator.

| Layer | Feature Maps | Filter | Stride | Input Size | Output Size |
|---|---|---|---|---|---|
| Input | - | - | - | 1*3120*1 | - |
| C1 | 10 | 120*1 | 5*1 | 1*3120*1 | 10*601*1 |
| P1 | 10 | 46*1 | 3*1 | 10*601*1 | 10*186*1 |
| C2 | 5 | 36*1 | 3*1 | 10*186*1 | 5*51*1 |
| P2 | 5 | 24*1 | 3*1 | 5*51*1 | 5*10*1 |
| FC | 5 | - | - | 5*10*1 | 25 |
| SOFTMAX | - | - | - | 25 | 25 |
| OUTPUT | - | - | - | 25 | 1 |

Figure 8. Layer's Parameters

### *3. MODEL EVALUATION*

To assess the system's execution, we have employed a variety of metrics that include precision, recall, F1-score, and model accuracy. The Pearson correlation formula is adopted to determine the model efficiency with the correlation measure of the train and test data graph. The preceding three metrics precision, recall and F1-score, are calculated with the cooperation of the confusion matrix.

1. Precision: It is the ratio between the accurately prognosticated true positives (TP) against the sumof both true positives (TP) and false positives (FP).

$$\text{Precision} = \frac{\text{TP}}{(\text{TP} + \text{FP})}.$$

2. Recall: It is the ratio between accurately predicted true positives (TP) upon the sum of both true positives (TP) and false negatives (FN).

$$\text{Recall} = \frac{\text{TP}}{(\text{TP} + \text{FN})}.$$

3. F1-Score: It is the hybrid standard of both the precision and recall weighted averages. F1-score scheming takes both the values of FP and FN to find the model performance as an alternative to the accuracy.

$$\text{F1-Score} = 2 * \frac{(\text{Precision} * \text{Recall})}{(\text{Precision} + \text{Recall})}.$$

4. Similarity measure: The model efficiency for validating the training and testing data for each epoch is plotted in Figure 9. such that we can assume that the model which has been designed is operating accurately with precise results. The correlation measure between train and test data graph has been determined employing the Pearson correlation [33] variable formula given below, where n = number of samples, x = train, and y = test.

$$r = \frac{n\left(\sum xy\right) - \left(\sum x\right)\left(\sum y\right)}{\sqrt{\left[n\sum x^2 - \left(\sum x\right)^2\right]\left[n\sum y^2 - \left(\sum y\right)^2\right]}}.$$

5. Selection: There are multiple methods to distinguish various classes of stocks according to the preferences of the investors; some of them prefer lengthy-term investment, while others present more attention in short term investment, also known as speculative investments. It is normal to observe in the stock report of a company that is reporting an ordinary performance; the stock price increasing drastically; this is one of the common events that illustrate that the stock price prognostication has no fixed rule, thus find effective features before train the data model is required. This proposal is focused on the short term price trend prediction, which makes this labour even more complex.

We have to determine the data with no labels; the initial action is to label the data. Identification should be added to the price trend by examining the actual closing price with the closing price of n trading days ago, the range of n is between 1 to 10 considering that this research is focused on the short term predictions. If the price trend increase, it has to be identified as 1 or mark as 0 in the opposing scenario. For purposes of optimization, the indices were employed from the indices of n−1th day to predict the price trend of the n th day.

6. Dimensionality: It is required to filter the high-quality stocks to ensure the best execution of the prognostication model; the data first has to be evaluated as a primordial step. The recursive feature elimination (RFE) was implemented to guarantee that all the chosen characteristics are useful since the raw data contains a vast number of features. If we include all the features into consideration, it will drastically increase the computational complexity and additionally produce side consequences for further research if we would like to perform fully unsupervised learning.

It was noticed that most of the previous works in the technical domain were examining all the stocks, while in the financial field, researchers favour the analysis of the specific scenario of investment for its practicality and look for results, to fill the gap between the two domains, It was decided to apply a feature extension based on the findings collected from the financial domain before the start the RFE procedure.

3. Forecasting: The plan of this approach is to model the data into time series to short-time forecasting in consequence, the larger number of the characteristics, the more complex the training procedure will be. So, to tackle this problem, the dimensionality reduction was implemented by using Randomized PCA at the origin of our suggested solution architecture.

***The codes referring to the proposed approach will be publicly available, at GITHUB under the user name WilfredoTovar.***

## VI. RESULTS AND EVALUATION

The evaluation of the algorithm was produced on a Hp laptop with 2.2 GHz Intel Core i7 processor, embedded 16 GB RAM memory. It was implemented by using Python 3.8.5, with the installed assortment of PyTorch and NumPy. Contrasted to the unvarying program, the identified neural network in PyTorch is active. The outcome of the experiment is then displayed using a variety of visual tool that supports PyTorch and NumPy. Others tools used were MatLab and R.

Representation of Data: We useddata obtained from the Yahoo financial website (https://finance.yahoo.com/). This dataset consists of 2954 stocks. We have downloaded 12 individual records of stocks for training. Each data file contained Financial data data detailed in different time frames.

1. Training Results: First, we contrasted the GAN with RNN-AE and RNN-VAE. All of the designs were prepared for 500 epochs practicing a series of around 3000 data points, a mini-bathc-size of around 100, and a learning rate of 10-−5. The loss of the GAN was determined with the equation in Figure 1. and the loss of RNN-AE was estimated as:

$$\max_{\theta} = \frac{1}{N}\sum_{i=1}^{N}\log p_{\theta}(y_i|x_i),$$

Where θ is the compounded set of parameters, N is the length of the financial serie, x-i is the i-th point in the series, which is the input of for the encoder, and y-i is the it-h spot in the series, which is the output from the decoder. The loss of RNN-VAE was determined as:

$$\sum_{i=1}^{N} L(\theta,\ \phi{:}\ x_i) = \sum_{i=1}^{N} -KL(q_{\phi}(\vec{z}|x_i)\|p_{\theta}(\vec{z}) + E_{q_{\phi}(\vec{z}|x_i)}[\log p_{\theta}(x_i|\vec{z})],$$

where → P θ ( z ) is usually a standard prior N ~ (0, 1), →Q φ ( z x) is the encoder, → θ p (x z ) is the decoder, and θ and φ are the sets of parameters for the decoder and encoder, respectively.

It was lengthened the RNN-AE to LSTM-AE and RNN-VAE to LSTM-VAE, and then contrasted the differences in the loss values of our design with these various generative models. Figure 9 shows the training effects, where the loss of the proposed GAN prototype was the lowest in the initial epoch, whereas all of the losses of the other designs were more than 20. Following 200 epochs of training, our GAN standard met to zero, while other designs simply commenced converging. At every step, the value of the loss function of the GAN was constantly significantly smaller than the losses of the other models.

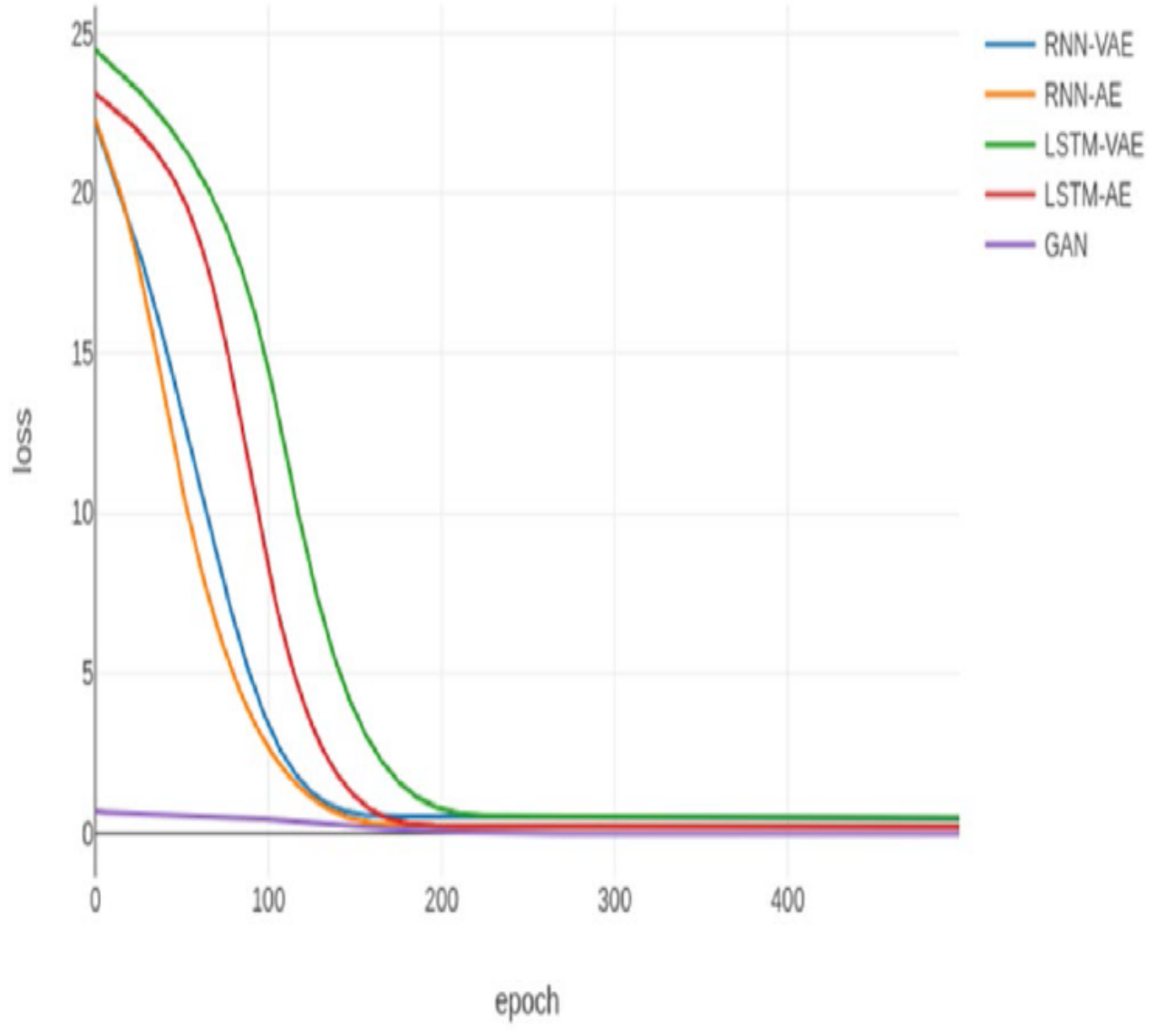

Figure 9. Losses of different generative models.

We then analyzed the outcomes acquired by the GAN models with those employing an LSTM, MLP (Multi-Layer Perceptron), CNN, and GRU as discriminators, that were labelled as BiLSTM-LSTM, BiLSTM-GRU, BiLSTM-MLP, and BiLSTM-CNN. Every design was trained with (500-epochs) with a (batch size-100), where the dimension of the chain involved a series of 3120 Financial-data points, and the learning rate was 1 × 10−5. Figure 10. illustrates the losses determined by GAN discriminators.

Figure 10. It is illustrated the loss by the MLP discriminator remained insignificant in the opening epoch and expanded after training for 200 epochs. The loss with the discriminator in the proposed model did insignificantly higher that with the MLP discriminator initially. However, it was perceptibly less than those of the GRU and LSTM discriminators. Finally, the loss converged quickly to zero with the proposed model, and it delivered the best of the analyzed models.

**Financial Data Generation:** Finally, it was implemented the designs gathered after training to produce Financial Data of Series Prediction by applying the GAN with the CNN, LSTM, MLP, and GRU as discriminators. The faint during the noise or volatile data points was established to 5, and the extension of the generated FinData was 400. Figure 11, exhibits the Fintech Data generated with different GANs; this data was set at a scale of variation per minute to try capture data to feed even the more complex strategies of investment such as scalping with high volatility usually per second. Figure 11, also proves that the Financial Data produced by the proposed prototype was more reliable in terms of the financial graphs morphology for high volatility stocks or even currencies. It was determined that despite the abundance of time actions, the Financial curvatures in the graphs produced employing the others models were distorted up when their models were exposed to high volatilities generations such the required for scalping analysis or intraday trading or at the commencement and end stages of the series, whereas the Financial data formed with the proposed design were not affected by this problem.

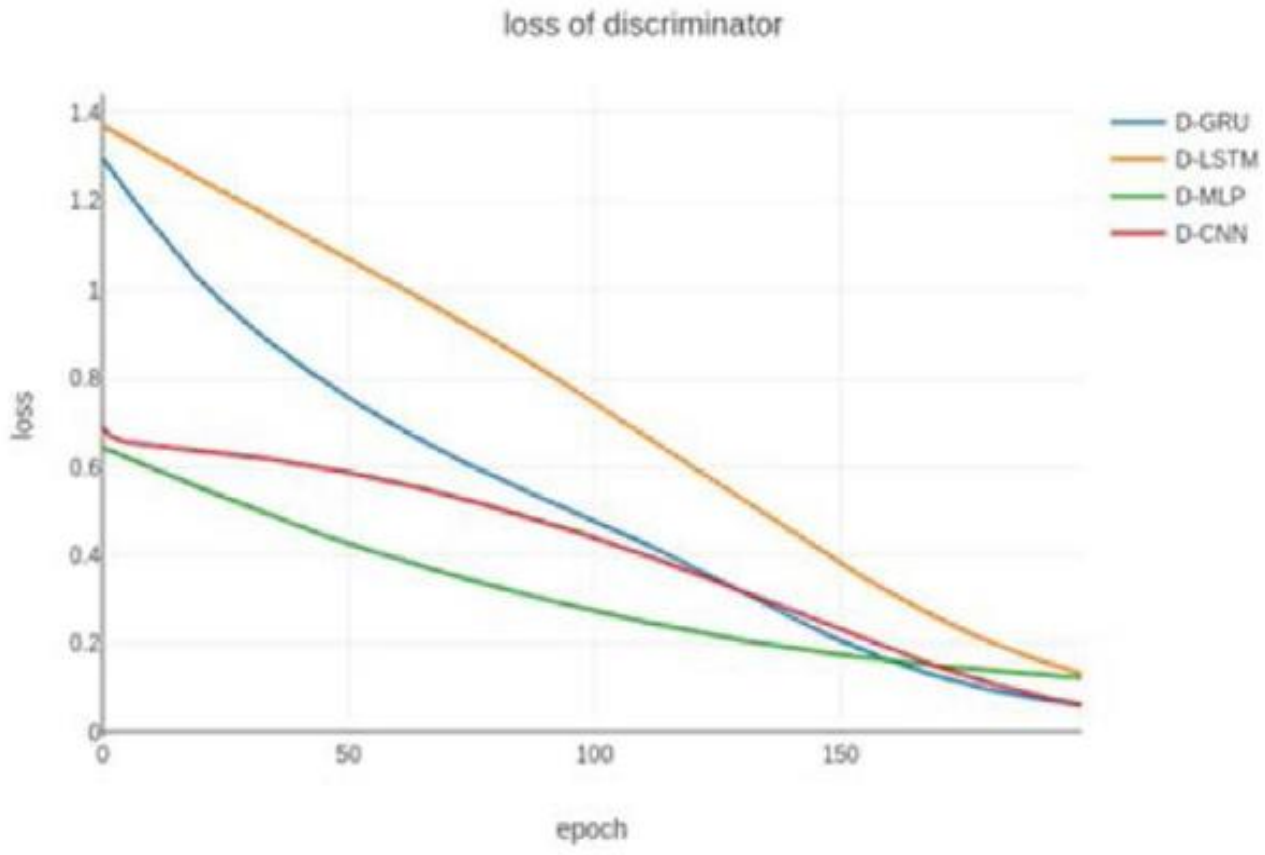

(a)

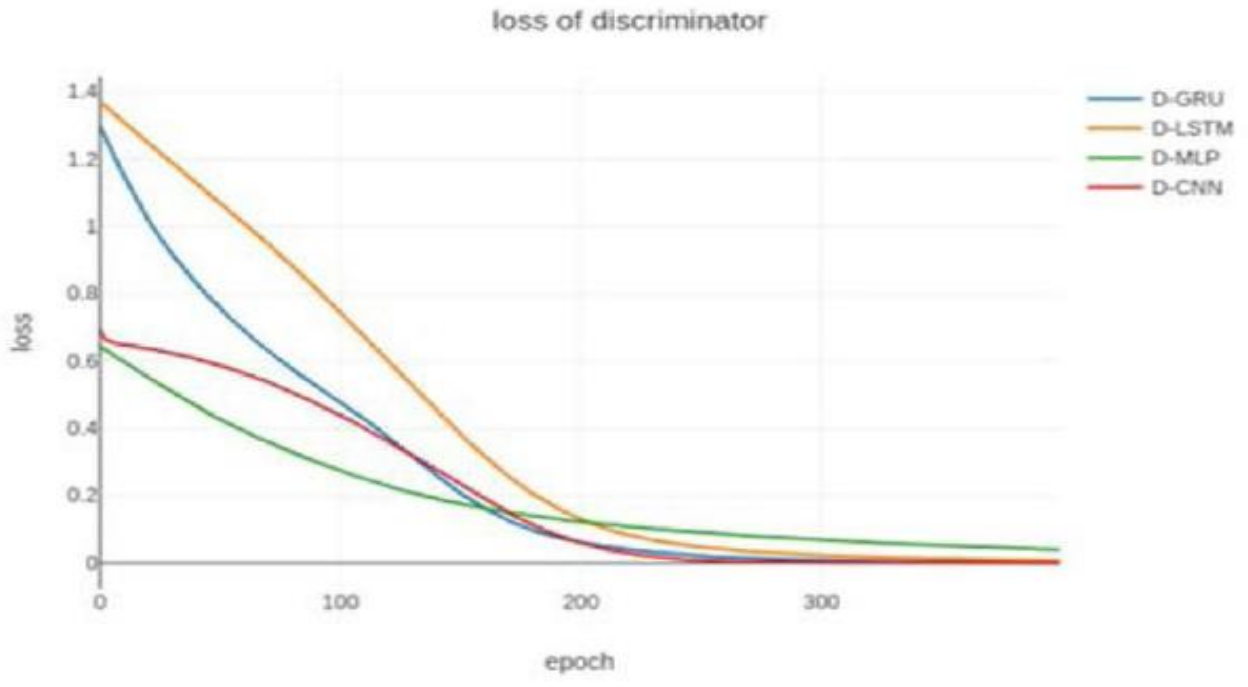

(b)

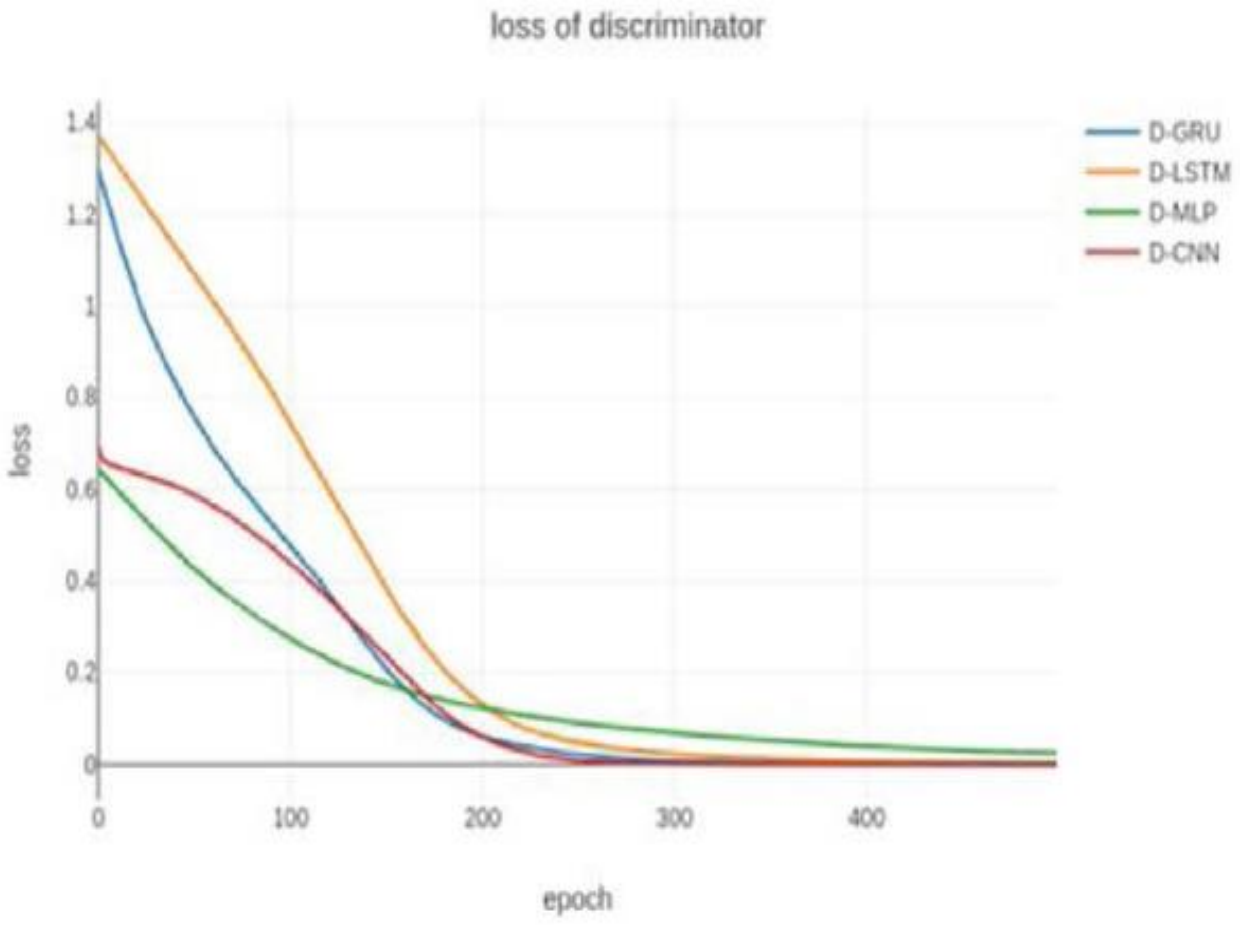

(c)

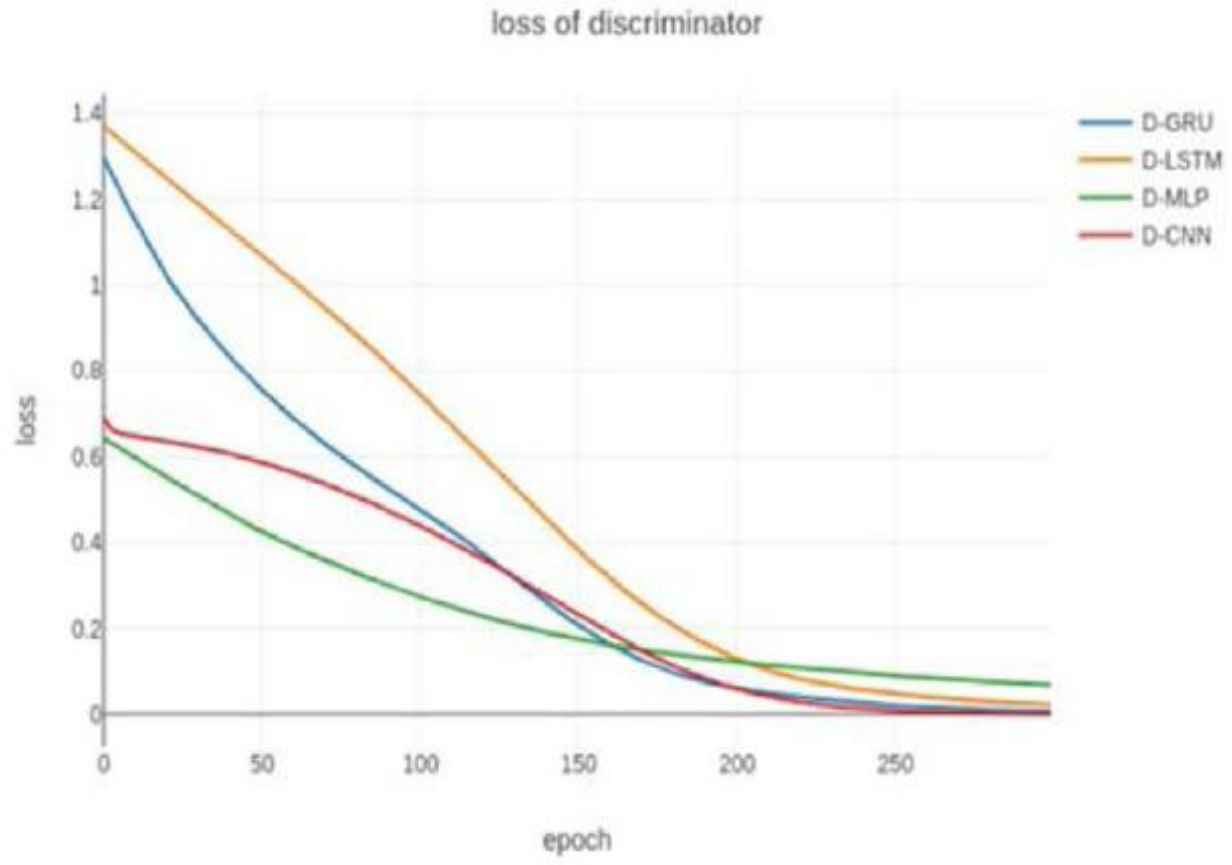

(d)

Figure 10. Loss of per class of discriminator. LSTM (orange line) , MLP (green line), CNN (red line) and GRU (blue line). (a–d) Represent the events after different epochs of training.

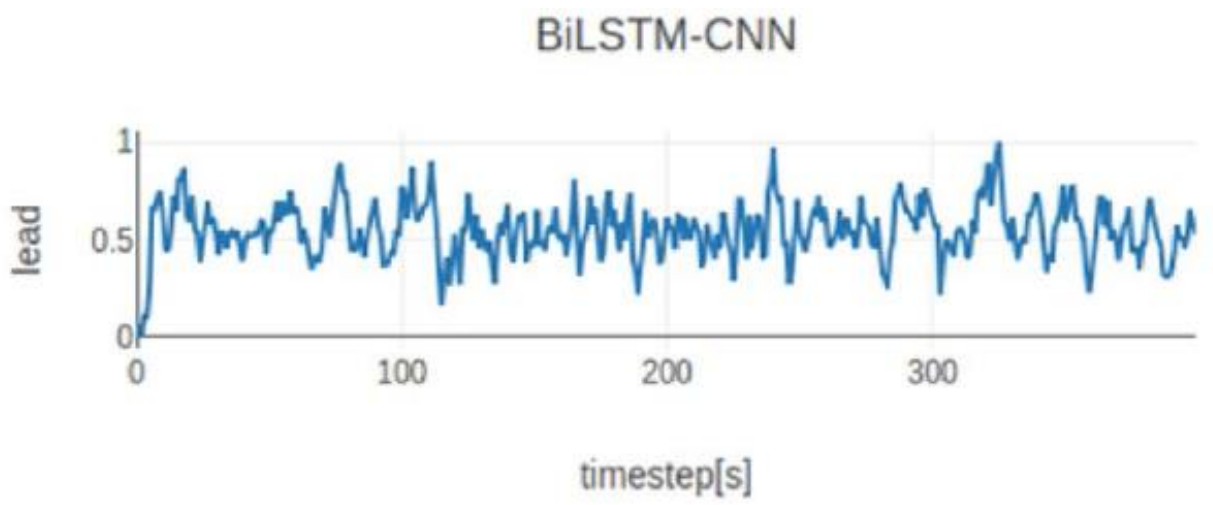

loss of discriminator

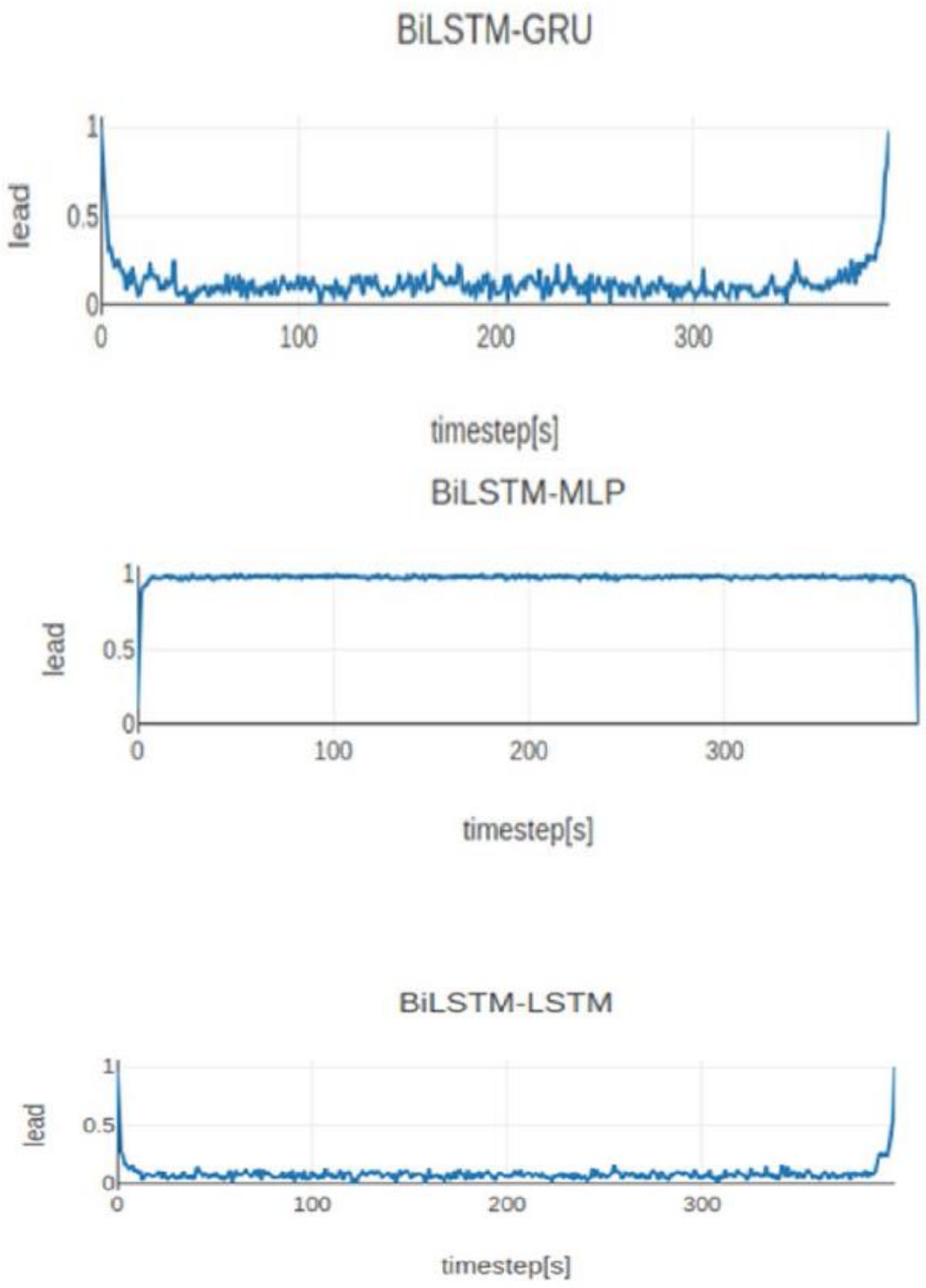

Figure 11. Results produced employing various discriminator structures.

| Method | PRD | RMSE | FD |
|---|---|---|---|
| BILSTM-CNN GAN | **66.408** | **0.276** | **0.756** |
| RNN-AE GAN | 121.877 | 0.506 | 0.969 |
| LSTM-AE GAN | 148.650 | 0.618 | 0.996 |
| RNN-VAE GAN | 146.566 | 0.609 | 0.982 |
| LSTM-VAE GAN | 145.978 | 0.607 | 0.975 |

Figure 12. Results of metrics' assessment for various generative models.

| Method | PRD | RMSE | FD |
|---|---|---|---|
| BiLSTM-CNN GAN | **51.799** | **0.215** | **0.803** |
| BiLSTM-GRU | 74.047 | 0.308 | 0.853 |
| BiLSTM-LSTM | 84.795 | 0.352 | 0.901 |
| BiLSTM-MLP | 147.732 | 0.614 | 0.989 |

Figure 13. Results of metrics' assessment of GANs with various discriminators.

**Percent root mean square difference (PRD)**[34]: is the extensively employed distortion estimation technique.

$$PRD = \sqrt{\frac{\sum_{n=1}^{N}(x_{[n]} - \widehat{x_{[n]}})^2}{\sum_{n=1}^{N}(x_{[n]})^2}} \times 100,$$

**The root mean square error (RMSE)** [34]**:** It indicates the steadiness among the original data and produced data, and it is estimated as:

$$RMSE = \sqrt{\frac{1}{N}\sum_{n=1}^{N}(x_{[n]} - \widehat{x_{[n]}})^2}.$$

**The Fréchet distance (FD)**[35]: It is a degree of correlation among curves that considers the position and order of points onward the curves, particularly in the event of time series data. A moderate or lower FD regularly holds higher quality and variety of generated results.

$$FD(P, Q) = \min \{||d||\}$$

Figure 12. shows that the proposed design possesses the least metric assessments regarding PRD, RMSE and FD contrasted with other generative models. It can be concluded that our model is the most reliable in producing Financial Time Series of Data compared with various alternatives of the autoencoder.

Figure 13. Illustrates that our suggested design achieved the best performance in terms of the RMSE, PRD and FD assessment confronted with various GANs. Additionally, it demonstrates that the Time Series Financial Data gathered employing the proposed

model was substantially similar to the Real-Time Financial Data from the stock market in terms of their tendencies Figure 14. Also, the GRU and LSTM are both adaptations of RNN, so their RMSE and PRD values were quite similar.

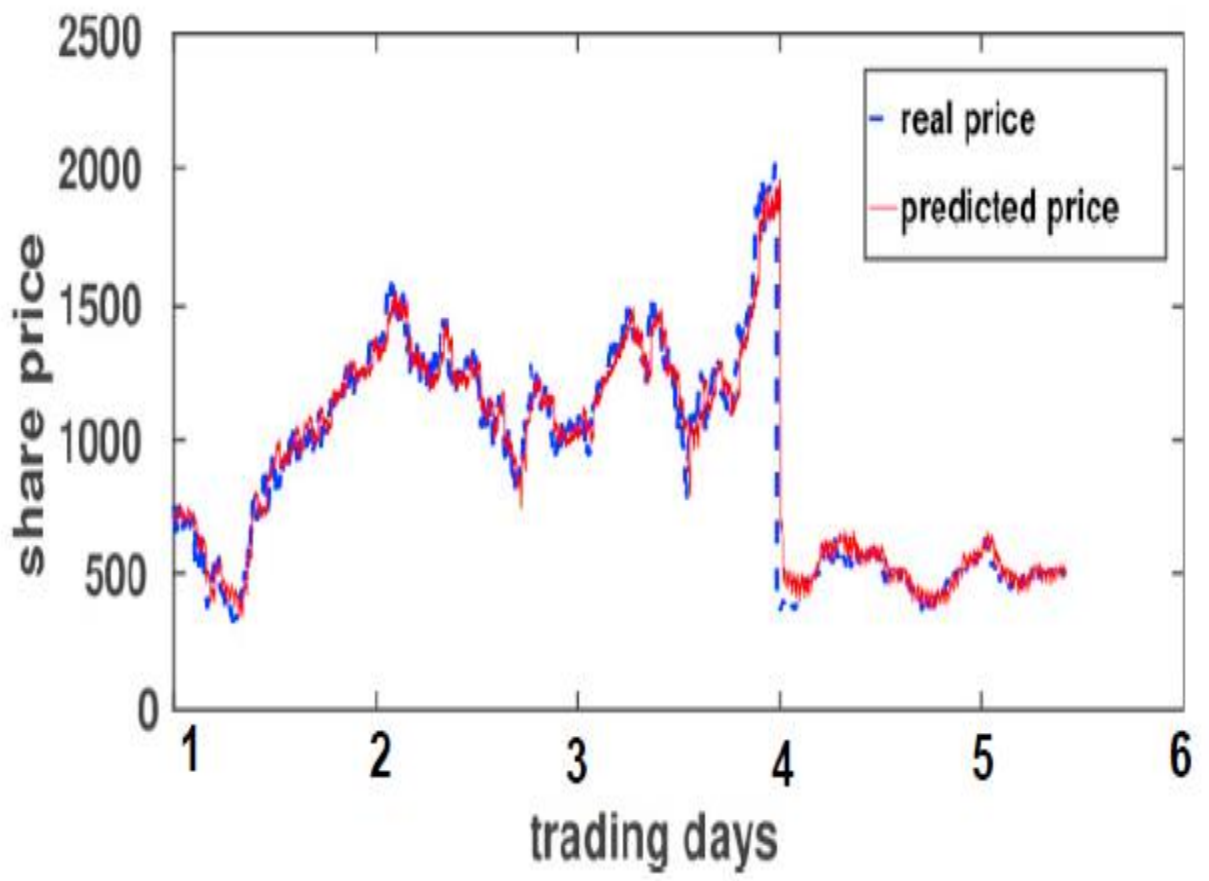

Figure 14. Short-time prediction assertiveness.

**The results of the Bi-LSTM-CNN network**: In this segment, the experimental outcomes of our proposed design have been presented, applying the different classifiers along with their feature importance graph and highest precision obtained. The results of the **Bi-LSTM-CNN** network have been evaluated in the form of the performance measure. Also, the traditional LSTM network approach results have been shown below for contrasting purposes with the list of all features given as an input to the recurrent neural network for training the model. In the performed analysis has been exposed to three companies results from three distinct stock exchanges named APPLE (NASDAQ: AAPL) from the Nasdaq Composite, AIR CHINA (SHA: 601111) from the Shanghai Composite Index, AIR CANADA (TSE: AC) from the Toronto Stock Exchange.

It has been demonstrated that the traditional design of the LSTM network is best suited for the stock market forecast, and it is recognized as a benchmark classifier employing the recurrent neural network by many of the researchers nowadays. Consequently, the traditional LSTM network has been generated without choosing the informational feature subset to train the model. It can be observed in Figure 15. This figure unquestionably exhibits the uncertain model behaviour with regard to the training and examining dataset. By observing these charts, it is obvious that the design has not been trained with the most relevant features, which caused a degradation of the model performance, deserts to determine the pattern in the dataset, and continuously presents the high variation in the model accuracy which, make it unsuitable for financial forecasting. So to overcome this constraint, it has been introduced our hybrid proposal to improve the predictive ability and precise model development. Further, to exclude the irrelevant features Wrapper method with the mentioned in previous sections, machine learning techniques.

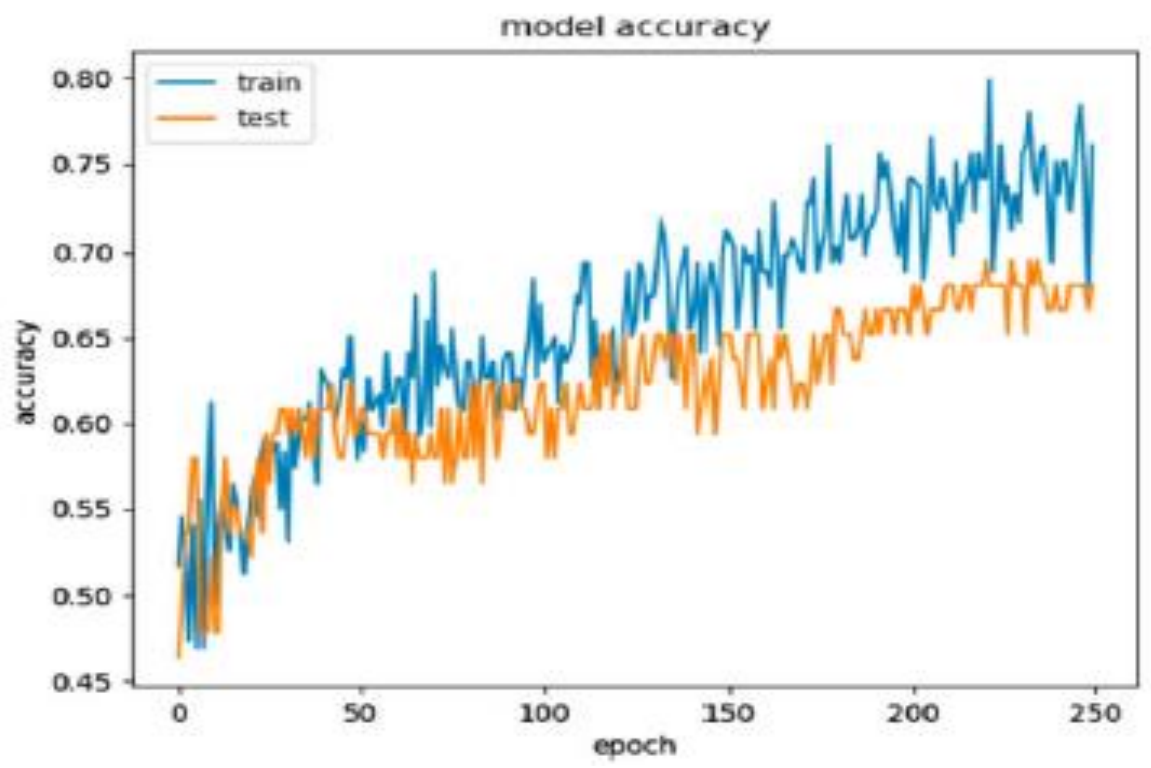

a) APPLE (NASDAQ: AAPL)

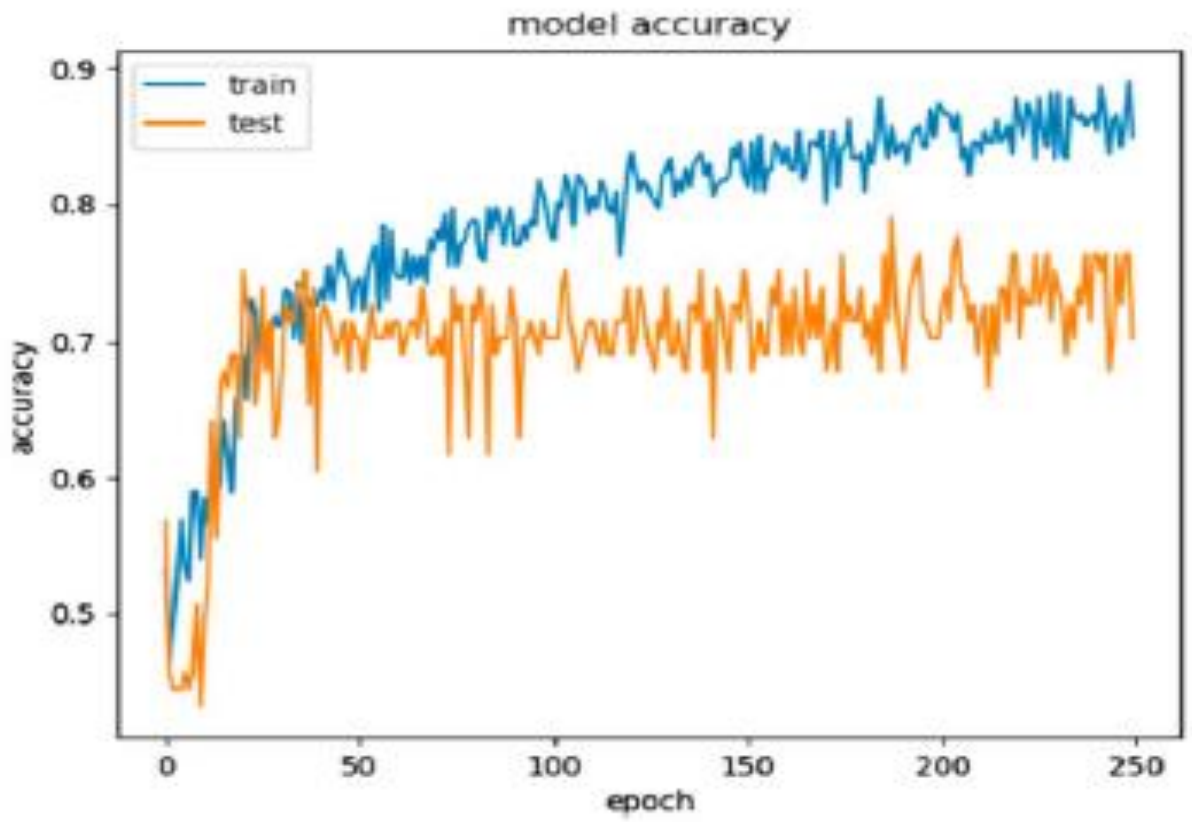

b) AIR CHINA (SHA: 601111)

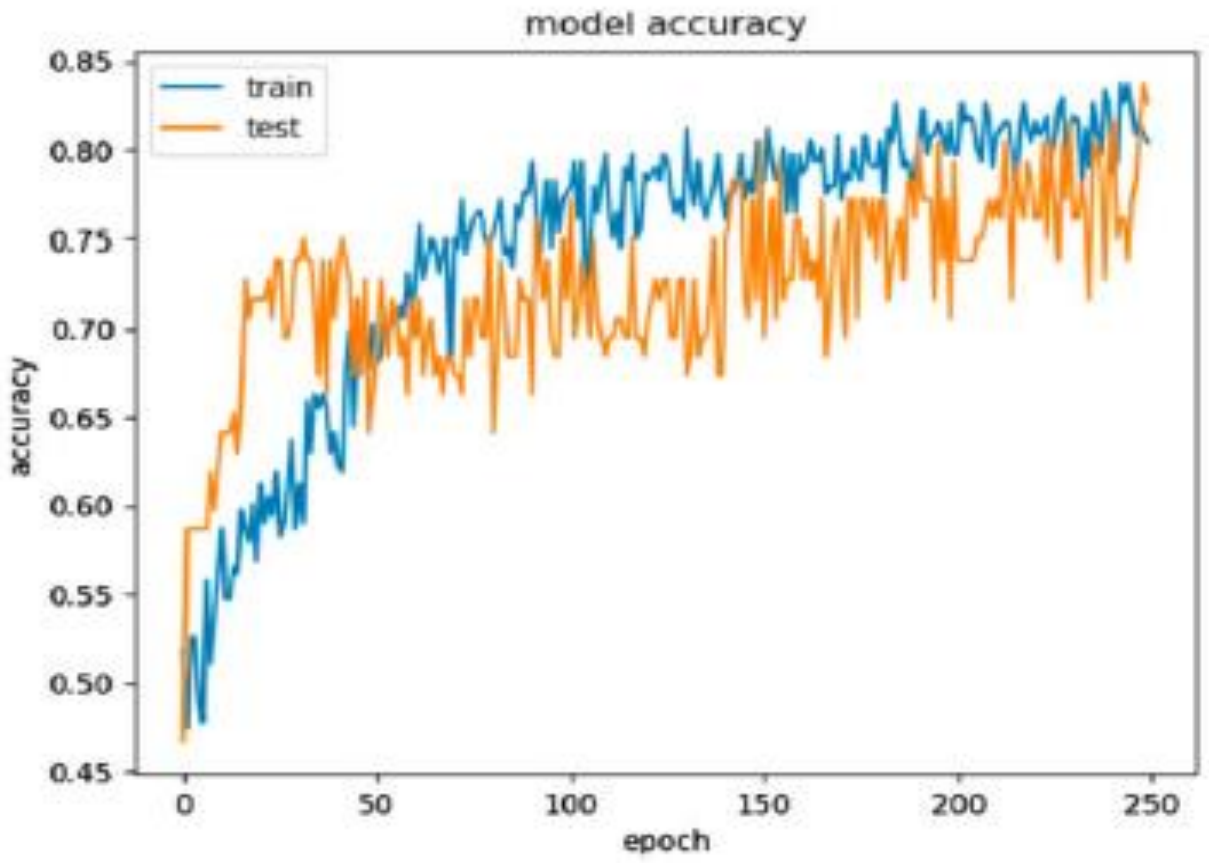

c) AIR CANADA (TSE: AC)

Figure 15. The plot of model precision employing the LSTM network for train and test data without performing the hybrid classifier process.

After selecting the best performing feature subset dynamically, using our approach with Bi-LSTM as the generator and the CNN as the discriminator. Figure 16. Shows the pattern of the model accuracy on both the dataset that is training and testing dataset. The conclusion can be extracted from Figure 16 where the precise model has been developed, which behaves in a more effective way with the high level of classification contrasted with a single classifier on LSTM network.

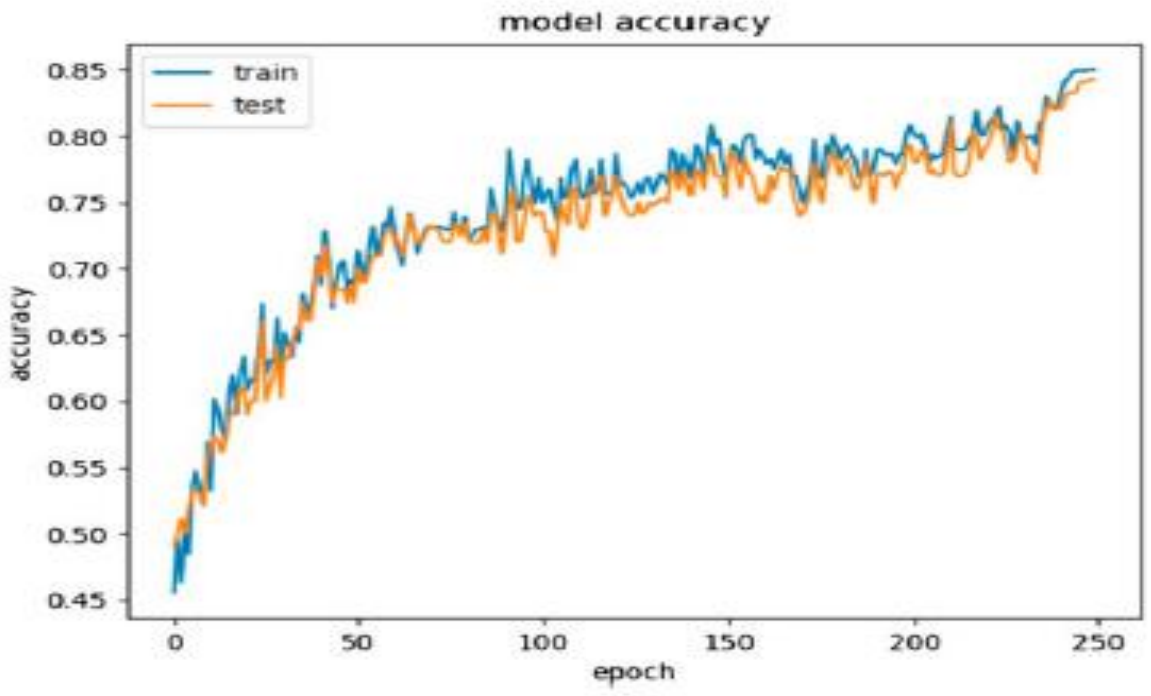

a) APPLE (NASDAQ: AAPL)

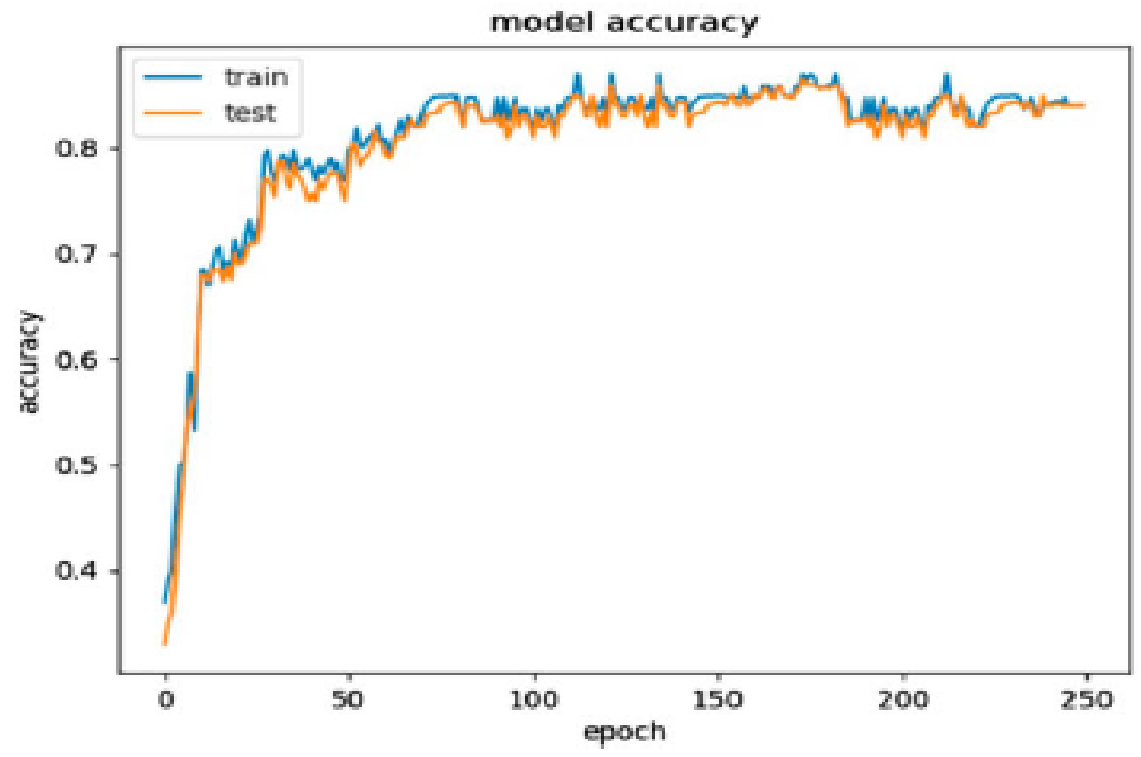

b) AIR CHINA (SHA: 601111)

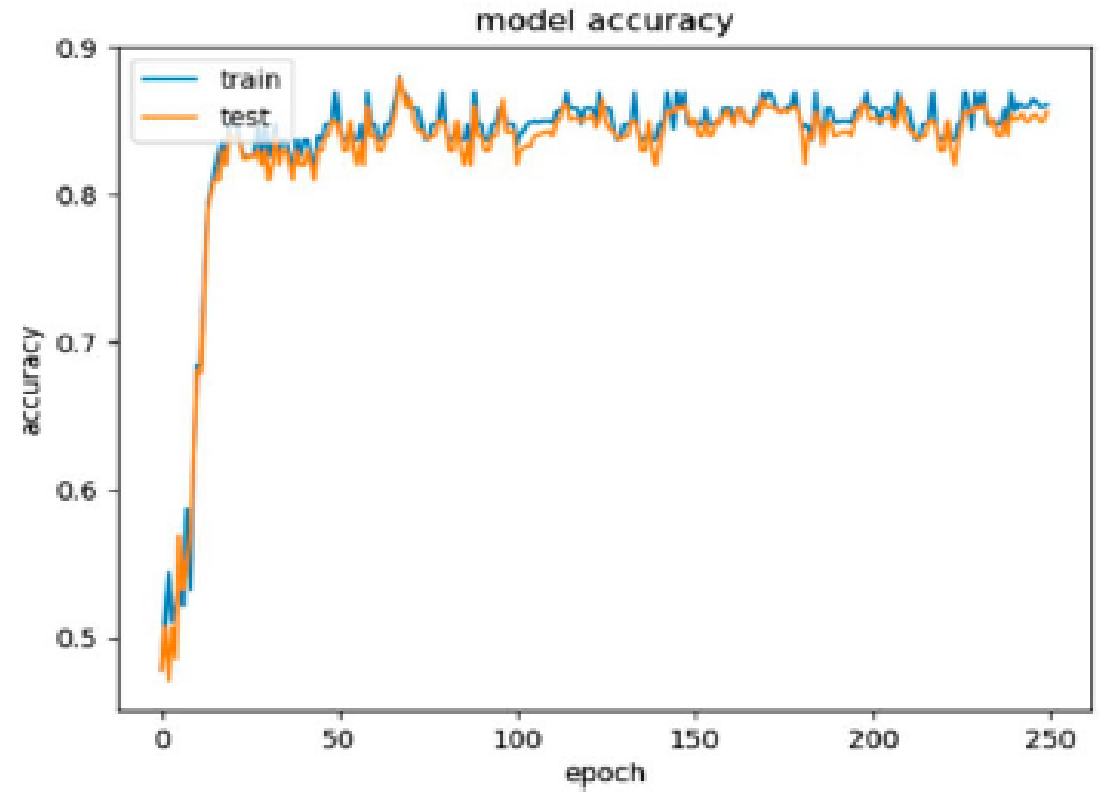

c) AIR CANADA (TSE: AC)

Figure 16. The plot of model precision employing the proposed Bi-LSTM-CNN network performing the hybrid classifier process.

| Approach | Feature selection | Precision | Recall | F1-score |
|---|---|---|---|---|
| LSTM | No | 0.54 | 0.57 | 0.55 |
| RFE-MLRC | Yes | 0.53 | 0.54 | 0.53 |
| RFE-DT | Yes | 0.53 | 0.53 | 0.53 |
| RFE-SVM | Yes | 0.55 | 0.56 | 0.55 |
| (RFE-MLRC) + LSTM | Yes | 0.60 | 0.65 | 0.62 |
| (RFE-DT) + LSTM | Yes | 0.61 | 0.62 | 0.62 |
| **Bi-LSTM-CNN** | **Yes** | **0.64** | **0.66** | **0.64** |

Figure 17. Comparative results with other traditional LSTM approaches with multiple and single classification.

The results of the parallel study exhibited in Figure 17. exposes the effectiveness of the proposed approach over its traditional and compounded pairs. As a result of a single classifier approach without performing feature selection using the LSTM network has achieved the minimum average reaching a maximum accuracy of 55% where precision = 0.54, recall = 0.57, and F1-score = 0.55. On the other hand, Our experimental outcomes also conclude that among different classifiers (RFE-SVM) with the LSTM network, the proposed approach BI-LSTM-CNN showed the maximum level of accuracy shows high accuracy with an average accuracy of 64% where precision = 0.64, recall = 0.66, and F1-score = 0.64 which outperforms the existing model.

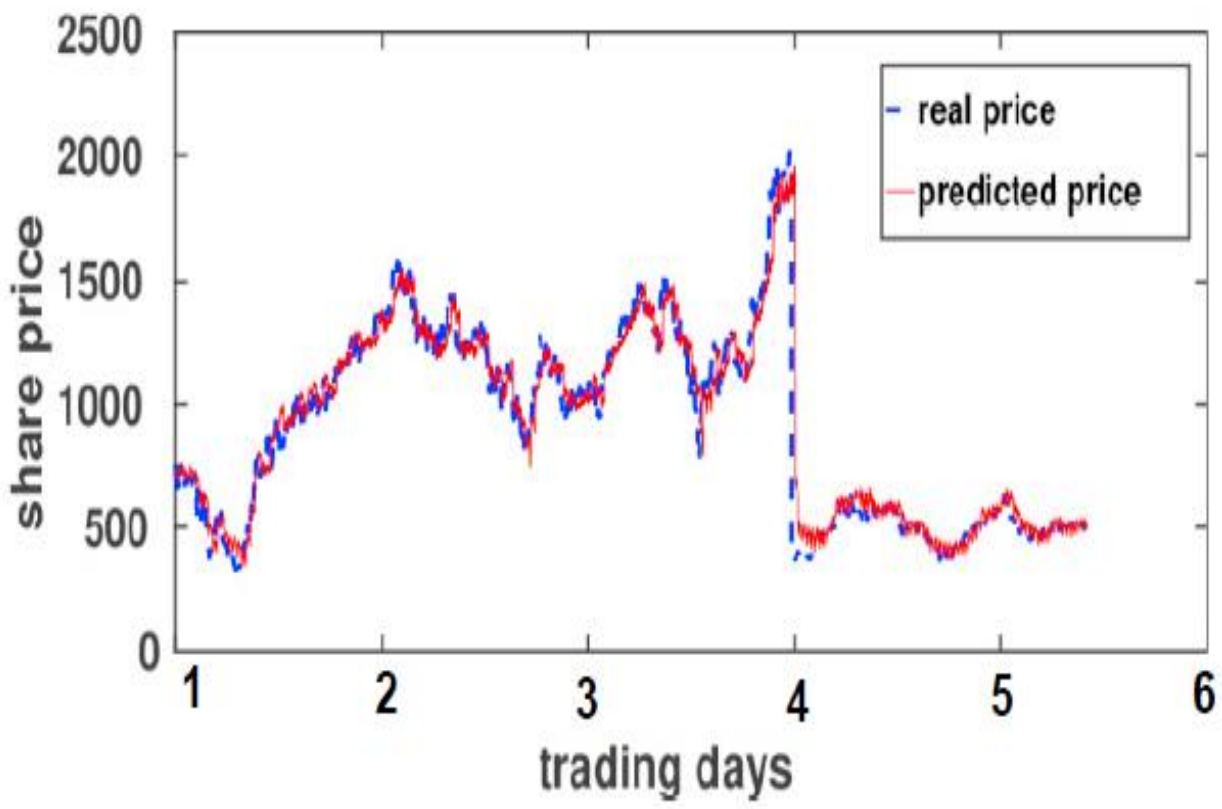

Figure 18. Short-time prediction assertiveness.

In Figure 18. The performance of the traditional LSTM approach in short-term predictions is illustrated on the left of the graph. It can be observed that the approach is capable of identifying price trends effectively, which for speculative transactions makes it is effective since they required a higher degree of effectiveness on price prediction, not only trends.

.

## VII. FUTURE WORK AND RESEARCH OPPORTUNITY

The research possibility is ample considering it was managed a mixture of data; Firstly, the regular trading data is available, researchers may employ the price information to determine largest of the technical indices, also they can be modelled with necessary prices in a determine range of time to apply time series and obtain the price or trend prediction also known as "Time Series Predictions." With a higher precision rate. For this objective, the implementation of Deep Learning techniques based on Transcoding Selection in [42] proven successfully for the stochastic optimization problem may be suitable for the financial field since it achieves high transcoding revenue while meeting the quality of service (QoS) requirements, and it can well handle dynamic cases[42]. Secondly, the value and indices may be utilized as characteristics, and joined with other data assembled in the dataset has the potential used as characteristics that may be employed for data mining purposes. Subsequently, another possibility of improvement refers to leverage the news scouring on media to observe their effects and e-posting on the stock market price, which eventually explore a real-time sentiment analysis system.

There are others challenges to overcome for time series prediction in the financial sector as is the scalability of the systems due to the complex requirements of computability and information processing; for example, in North America, more than 10,000 stocks are handled daily, which generate data every second of operation causing the impossibility of effective global analysis. However, a Deep Learning approach could be adopted and transferred to the financial domain as a framework solution presented in [43][44] for Resource Optimization with Edge Computing, which suggests a joint optimization framework about caching. Additionally, to increase the level of security and privacy of information in centralized devices, Blockchain notions shown in [45] can be adapted, which can be customizable due to its approach to the integration of software-defined networking (SDN) and different devices. Finally, there are models and concepts of deep reinforcement learning that have achieved remarkable results in other disciplines such as in [46],[47],[48], for which their potential applicability could be explored for the stock market and times series prediction.

On this paper have produced a satisfactory prognostication of price result from the proposed model. However, this research project still possesses considerable potential in future research. Primarily, the objective of building the model to produce time series price trends prediction is to complete the very first step of stock market price prediction. With a reliable trend prediction, we can perform the price prediction in a more reliable and scalable way for different magnitudes of times.

## VIII. CONCLUSION

This research paper offered multiple contributions. It contribute with the research on the lack of adequate Time Series Prediction and generative data for financial stock market research, It was generated a novel deep learning design that can produce Financial Time Series Data from public markets data without missing the peculiarities of the existing data. The design is based on the GAN, where the Bi-LSTM is employed as the generator, and the CNN is utilized as the discriminator. After training with Stock Market Data, our design can produce synthetic Stock Market Data points and configure models of assertive market tendencies that match the data patterns in the original Financial data. The proposed model achieved a more reliable performance than the other two deep learning designs in both the preparation and testing stages, and it was advantageous confronted with three other generative designs at producing financial time-series data.

The Financial data integrated applying to the proposed model were morphologically similar to the real financial tendencies. The evaluation results validate the effectiveness of the application of concepts of machine learning, which are very helpful in the neural network while developing the LSTM network. This maximizes the predictive capability by choosing the most beneficial performing feature subset given as an input to the LSTM network which trains the model with the most appropriate characteristics, increases the interpretability by adjusting the exchange behaviour that is already present in the dataset and discarding the unnecessary characteristics, reduces the complexity and decreased the training time of the design with high accuracy.

## REFERENCES

[1] Krollner, Bjoern, Bruce Vanstone, and Gavin Finnie. "Financial time series forecasting with machine learning techniques: A survey." (2010).

[2] Agrawal, J. G., V. S. Chourasia, and A. K. Mittra. "State-of-the-art in stock prediction techniques." International Journal of Advanced Research in Electrical, Electronics and Instrumentation Engineering 2.4 (2013): 1360-1366.

[3] Saad, Emad W., Danil V. Prokhorov, and Donald C. Wunsch. "Comparative study of stock trend prediction using time delay, recurrent and probabilistic neural networks." Neural Networks, IEEE Transactions on 9.6 (1998): 1456-1470.

[4] Rather, Akhter Mohiuddin, Arun Agarwal, and V. N. Sastry. "Recurrent neural network and a hybrid model for prediction of stock returns." Expert Systems with Applications 42.6 (2015): 3234-3241.

[5] Kim, K. J. (2003). Financial time series forecasting using support vector machines. Neurocomputing, 55(1-2), 307-319

[6] Nair, B. B., Mohandas, V. P., & Sakthivel, N. R. (2010). A decision tree—rough set hybrid system for stock market trend prediction. International Journal of Computer Applications, 6(9), 1-6.

[7] Chen, K., Zhou, Y., & Dai, F. (2015, October). A LSTM-based method for stock returns prediction: A case study of China stock market. In 2015 IEEE international conference on big data (big data) (pp. 2823-2824). IEEE.

[8] Chong, E., Han, C., & Park, F. C. (2017). Deep learning networks for stock market analysis and prediction: Methodology, data representations, and case studies. Expert Systems with Applications, 83, 187-205.

[9] Cho, K., Van Merriënboer, B., Gulcehre, C., Bahdanau, D., Bougares, F., Schwenk, H., & Bengio, Y. (2014). Learning phrase representations using RNN encoder-decoder for statistical machine translation. arXiv preprint arXiv:1406.1078.

[10] Kingma, D. P., & Welling, M. (2013). Auto-encoding variational bayes. arXiv preprint arXiv:1312.6114.

[11] K. Kim, "Financial time series forecasting using support vector machines," Neurocomputing, vol. 55, no. 1–2, pp. 307 – 319, 2003, support Vector Machines. [Online]. Available: http://www.sciencedirect.com/science/article/pii/S0925231203003722.

[12] B. Batres-Estrada, "Deep learning for multivariate financial time series," 2015.

[13] A. Sharang and C. Rao, "Using machine learning for medium frequency derivative portfolio trading," CoRR, vol. abs/1512.06228, 2015. [Online]. Available: http://arxiv.org/abs/1512.06228

[14] J. B. Heaton, N. G. Polson, and J. H. Witte, "Deep learning in finance," CoRR, vol. abs/1602.06561, 2016. [Online]. Available: http://arxiv.org/abs/1602.06561

[15] K. Greff, R. K. Srivastava, J. Koutn'ık, B. R. Steunebrink, and J. Schmidhuber, "LSTM: A search space odyssey," arXiv preprint arXiv:1503.04069, 2015.

[16] Lei, L. (2018). Wavelet neural network prediction method of stock price trend based on rough set attribute reduction. Applied Soft Computing, 62, 923-932.

[17] Gheyas IA, Smith LS. A novel neural network ensemble architecture for time series forecasting. Neurocomputing. 2011 Nov 1;74(18):3855–3864.

[18] Evans C, Pappas K, Xhafa F. Utilizing artificial neural networks and genetic algorithms to build an algo-trading model for intraday foreign exchange speculation. Math Comput Model. 2013 Sep 1;58(5–6):1249–1266.

[19] Nayak RK,Mishra D, Rath AK. A Naive SVM-KNN based stockmarket trend reversal analysis for Indian benchmark indices. Appl Soft Comput. 2015 Oct 1;35:670–680.

[20] Cavalcante RC, Oliveira AL. An autonomous trader agent for the stock market based on online sequential extreme learning machine ensemble. In Neural Networks (IJCNN), 2014 International Joint Conference on 2014 Jul 6 (pp. 1424–1431). IEEE.

[21] Ballings M, Van den Poel D, Hespeels N, et al. Evaluating multiple classifiers for stock price direction prediction. Expert Syst Appl. 2015 Nov 15;42(20):7046–7056.

[22] McNally, S., Roche, J., & Caton, S. (2018, March). Predicting the price of bitcoin using machine learning. In 2018 26th Euromicro International Conference on Parallel, Distributed and Network-based Processing (PDP) (pp. 339-343). IEEE.

[23] Fischer, T., & Krauss, C. (2018). Deep learning with long short-term memory networks for financial market predictions. European Journal of Operational Research, 270(2), 654-669.

[24] Press, O., Bar, A., Bogin, B., Berant, J., & Wolf, L. (2017). Language generation with RGAN without pre-training. arXiv preprint arXiv:1706.01399.

[25] Li, J., Monroe, W., Shi, T., Jean, S., Ritter, A., & Jurafsky, D. (2017). Adversarial learning for neural dialogue generation. arXiv preprint arXiv:1701.06547.

[26] Hüsken, M. & Stagge, P. Recurrent neural networks for time series classification. Neurocomputing 50, 223–235, https://doi. org/10.1016/S0925-2312(01)00706-8 (2003).

[27] Bowman, S. R., Angeli, G., Potts, C., & Manning, C. D. (2015). A large annotated corpus for learning natural language inference. arXiv preprint arXiv:1508.05326.

[28] Zabalza, J., Ren, J., Zheng, J., Zhao, H., Qing, C., Yang, Z., ... & Marshall, S. (2016). Novel segmented stacked autoencoder for effective dimensionality reduction and feature extraction in hyperspectral imaging. Neurocomputing, 185, 1-10.

[29] Efimov, D., Xu, D., Kong, L., Nefedov, A., & Anandakrishnan, A. (2020). Using generative adversarial networks to synthesize artificial financial datasets. arXiv preprint arXiv:2002.02271.

[30] Binkowski, M., Marti, G., & Donnat, P. (2018, July). Autoregressive convolutional neural networks for asynchronous time series. In International Conference on Machine Learning (pp. 580-589).

[31] Schreyer, M., Sattarov, T., Reimer, B., & Borth, D. (2019). Adversarial learning of deepfakes in accounting. arXiv preprint arXiv:1910.03810.

[32] Hsu, C. M. (2013). A hybrid procedure with feature selection for resolving stock/futures price forecasting problems. Neural Computing and Applications, 22(3-4), 651-671.

[33] Byjus. Gdpr. [cited 2018 Aug 30]. https://byjus.com/pearsoncorrelation-formula 2018.

[34] Christoffersen, P., & Gonçalves, S. (2004). Estimation risk in financial risk management. CIRANO.

[35] Mittnik, S., & Rachev, S. T. (1991). Alternative multivariate stable distributions and their applications to financial modeling. In Stable processes and related topics (pp. 107-119). Birkhäuser Boston.

[36] Gorgulho, A., Gorgulho, A. M., Neves, R. F., & Horta, N. (2012). Intelligent financial portfolio composition based on evolutionary computation strategies. Springer Science & Business Media.

[37] Pang, X., Zhou, Y., Wang, P., Lin, W., & Chang, V. (2018). An innovative neural network approach for stock market prediction. The Journal of Supercomputing, 1-21.

[38] Nguyen, D. H. D., Tran, L. P., & Nguyen, V. (2019, November). Predicting Stock Prices Using Dynamic LSTM Models. In International Conference on Applied Informatics (pp. 199-212). Springer, Cham.

[39] Liu, Y. (2019). Novel volatility forecasting using deep learning–long short term memory recurrent neural networks. Expert Systems with Applications, 132.

[40] Yahoo. (2018). NIFTY 50 (NSEI)/S&P BSE SENSEX (BSESN). Retrieved April, 2020, from https://in.finance.yahoo.com

[41] Yahoo. (2019). S&P 500 Stock Data. Retrieved April, 2020, from https://finance.yahoo.com/quote/%5EGSPC/history?p=%5EGSPC.

[42] Liu, M., Teng, Y., Yu, F. R., Leung, V. C., & Song, M. (2020). A Deep Reinforcement Learning-based Transcoder Selection Framework for Blockchain-Enabled Wireless D2D Transcoding. IEEE Transactions on Communications.

[43] Li, M., Yu, F. R., Si, P., Wu, W., & Zhang, Y. (2020). Resource Optimization for Delay-Tolerant Data in Blockchain-Enabled IoT with Edge Computing: A Deep Reinforcement Learning Approach. IEEE Internet of Things Journal.

[44] Zhang, R., Yu, F. R., Liu, J., Xie, R., & Huang, T. (2020). Blockchain-incentivized D2D and mobile edge caching: A deep reinforcement learning approach. IEEE Network, 34(4), 150-157.

[45] Luo, J., Yu, F. R., Chen, Q., & Tang, L. (2020, June). Blockchain-Enabled Software-Defined Industrial Internet of Things with Deep Recurrent Q-Network. In ICC 2020-2020 IEEE International Conference on Communications (ICC) (pp. 1-6). IEEE.

[46] Du, J., Yu, F. R., Lu, G., Wang, J., Jiang, J., & Chu, X. (2020). Mec-assisted immersive VR video streaming over terahertz wireless networks: A deep reinforcement learning approach. IEEE Internet of Things Journal.

[47] Fu, F., Kang, Y., Zhang, Z., & Yu, F. R. (2020, July). Transcoding for Live Streaming-based on Vehicular Fog Computing: An Actor-Critic DRL Approach. In IEEE INFOCOM 2020-IEEE Conference on Computer Communications Workshops (INFOCOM WKSHPS) (pp. 1015-1020). IEEE.

[48] Fu, F., Kang, Y., Zhang, Z., Yu, F. R., & Wu, T. (2020). Soft Actor-Critic DRL for Live Transcoding and Streaming in Vehicular Fog Computing-Enabled IoV. IEEE Internet of Things Journal.